%% file: marquez.tex
\documentclass{aa}     
\usepackage{latexsym, epsfig}

\def\CEL{\textit{corrected Entropic Line}{}}
\def\EL{\textit{Entropic Line}{}}
\def\Se{S\'ersic{}}
\def\DV{de Vaucouleurs{}}
\def\diff{{\rm d}}
\thesaurus{02.07.1; 03.13.8; 11.03.4;
11.05.1; 11.05.2; 11.06.1; 11.06.2; 11.19.6}

\title{Gravo-thermal properties and formation of elliptical galaxies.
\thanks{Based on observations collected at the Canada France Hawaii
Telescope and at the European Southern Observatory, La Silla, Chile}}

\author {I. M\'arquez \inst{1}
\and
  G.B. Lima Neto \inst{2,3}
\and
  H. Capelato \inst{4}
\and
  F. Durret \inst{3,5}
\and
  D. Gerbal \inst{3,5}}

\offprints{I. M\'arquez, isabel@iaa.es}
\institute{
Instituto de Astrof\'\i sica de Andaluc\'\i a (CSIC),
Apartado 3004 , E-18080 Granada, Spain
\and
Instituto Astron\^omico e Geof\'{\i}sico/USP, Av. Miguel Stefano 4200,
S\~ao Paulo/SP, Brazil
\and
Institut d'Astrophysique de Paris, CNRS, 98bis Bd Arago,
F-75014 Paris, France
\and
Instituto de Pesquisas Espaciais, S\~{a}o Jos\'e dos Campos/SP, Brazil
\and
DAEC, Observatoire de Paris, Universit\'e Paris VII, CNRS (UA 173),
F-92195 Meudon Cedex, France
}
\date{Received, 1999; accepted }
\begin{document}

\maketitle

\begin{abstract}

We have analyzed a sample of galaxies belonging to three clusters: Coma, Abell
85, and Abell 496 (\textit{real galaxies}) and a sample of simulated
elliptical galaxies formed in a hierarchical merging scheme (\textit{virtual}
galaxies). We use the S\'ersic law to describe their light profile. The specific
entropy (Boltzmann-Gibbs definition) is then calculated supposing that the
galaxies behave as spherical, isotropic, one-component systems. We find that, to
a good approximation ($\sim 10$\%), both \textit{real} and \textit{virtual} galaxies have an
almost unique specific entropy. Within this approximation the galaxies are
distributed in a thin plane in the space defined by the three S\'ersic law
parameters, which we call the \textit{Entropic Plane}. A further analysis shows
that both \textit{real} and \textit{virtual} galaxies are in fact located on a
thin line, therefore indicating the existence of another -- and yet unknown --
physical property, besides the uniqueness of the specific entropy.

A more careful examination of the \textit{virtual} galaxies sample indicates a
very small increase of their specific entropy with merging generation. In a
hierarchical scenario, this implies a correlation between the specific entropy
and the total mass, which is indeed seen in our data. The scatter and tilt of
the \textit{Entropic Line}, defined by Lima Neto et al. (1999a), are reduced
when this correlation is taken into account. Although one cannot distinguish
between various generations for \textit{real} galaxies, the distribution of
their specific entropy is similar to that in the \textit{virtual} sample,
suggesting that hierarchical merging processes could be an important mechanism
in the building of elliptical galaxies.


\end{abstract}

\section{Introduction}

Elliptical galaxies present a striking regularity in their global
luminosity distributions. Within a wide range of sizes, the light
profile of elliptical galaxies can be described by a non-homologous
generalization of the de Vaucouleurs $R^{1/4}$ profile, the S\'ersic
law (e.g. Caon et al. 1993; Graham \& Colless 1997; Prugniel \& Simien
1997).

This regularity may be understood in terms of a relaxation process:
elliptical galaxies seem to be in a quasi-equilibrium state, implying
that they should obey the virial theorem. From the second law of
thermodynamics, a dynamical system in equilibrium is in a maximum
entropy state. Due to their peculiar properties (long range unshielded
interactions, equivalence of inertial and gravitational mass, etc.)
the thermodynamics of gravitational systems present some difficulties,
as well explained in academic books (e.g. Saslaw 1985).

For these systems an equilibrium state is never really
reached. Various dynamical time scales can be defined: the natural
\textit{dynamical} time $t_{\rm d} \approx 1/\sqrt{4 \pi G \rho}$
(where $\rho$ is the mean density of the system), the
\textit{violent relaxation} time scale, $t_{\rm VR}$, with $t_{\rm VR}
\approx t_{\rm d}$, related to the phase mixing process which leads to
a quasi-equilibrium state, and a large \textit{secular} time scale
$t_{\rm sec}$, which is related to the slow effects of two-body
gravitational interactions. It is essentially on this scale that one
can assert that the equilibrium of a self-gravitating
system is \textit{never} established.

For elliptical galaxies, we have $t_{\rm VR} \approx \varepsilon\,
t_{\rm sec}$ with $\varepsilon \approx (N/\log N)^{-1} \approx
10^{-8}$ (where $N$ is the number of particles). Therefore even if the
entropy $S$ of a galaxy is ever growing on the secular time scale,
after violent relaxation we have $\mathrm{d} S / \mathrm{d} t_{\rm VR}
\approx \varepsilon$. Stating that the system is in a
quasi-equilibrium stage is equivalent to saying that the entropy is
quasi-constant.

However, maximizing the entropy results in an isothermal sphere
(Lynden-Bell 1967) which is not valuable either from the point of view
of physics (divergent total mass) or from observations (observed
density profiles are steeper than the isothermal profile; see also
White \& Narayan 1987).

It is important to note that although there are no exact stationary
entropy states for self-gravitating systems (that is, no absolute
maximum entropy states), lowest energy states may exist, as suggested
by Wiechen et al. (1988). In order to reach such equilibrium states,
the system must necessarily undergo a violent relaxation phase, be it
through a collapse or a merger.  However the final configuration
reached depends, in principle, on how strong the violent relaxation
phase was. This raises the interesting question of how these
equilibria, based on minimum energy, would relate to the final entropy
of the system.

Numerous works have been devoted to the entropy problem (see for
instance Merritt 1999). In a previous paper (Lima Neto et al. 1999a,
hereafter LGM), a different approach has been adopted: instead of
trying to obtain the final expected configuration by maximizing the
entropy, LGM admit the existence of a state of quasi-constant entropy
and calculate this entropy by deriving it from the observed light
(mass) distribution. In order to compare objects of different masses,
LGM introduced the specific entropy $s = S/M$, that is the entropy
normalized by the mass. The specific entropy was then calculated by
assuming that the stars obey the equations of state of an ideal gas
and using the standard thermodynamical definition of the entropy. LGM
showed that the galaxies of two clusters and a group had the same
value of $s$ and, therefore, that one could derive relative distances
between these clusters using the S\'ersic profile to model the light
distribution. LGM suggested that the galaxies having an unique $s$
could explain distance indicators based on the shape of the brightness
profile of galaxies, like those proposed by Young \& Currie (1994,
1995).

As in LGM, we will describe the light distribution of an
elliptical galaxy using a S\'ersic profile:
\begin{equation}
\Sigma (R)=\Sigma _{0}\exp (-(R/a)^{\nu})
\label{sersic}
\end{equation}
characterized by three primary parameters:
$\nu$, the shape parameter (independent of cluster distance), $a$, the
scale parameter (distance dependent, in arcsec), and $\Sigma_{0}$, the
intensity parameter (in erg s$^{-1}$ arcsec$^{-2}$).

In contrast with LGM, who
used the thermodynamical definition of the entropy, we will adopt 
here the microscopic Boltzmann-Gibbs definition, therefore eliminating the 
assumption based on the equations of state of an ideal gas. Assuming that 
elliptical galaxies are well described by the S\'ersic law, 
we have derived the specific entropy (see details in Appendix I):
\begin{equation}
s(a, \nu, \Sigma_0) = 0.5\ln(\Sigma_0) + 2.5 \ln(a) + F(\nu),
\label{intro}
\end{equation}
with:
\begin{equation}
     F(\nu) \equiv + 0.2 \ln(\nu) - \frac{1.3}{\nu} + 3.9\nu^{-1.3} - 2.7.
     \label{entronu}
\end{equation}
%
Should the specific entropy of galaxies, $s(a, \nu, \Sigma_0)$, be a
constant or, at least, display a small dispersion around its mean
value, then Equation (\ref{intro}) would define a thin surface in the
parameter space [$\Sigma_0, a, \nu$], or a plane in the space
[$\ln(\Sigma_0), \ln(a), F(\nu)$]. The results presented in LGM
suggest that this is indeed the case.

In the next section we describe the data used in this paper, i.e.  the
surface brightness of cluster galaxies as well as that of simulated
galaxies. We also discuss the fitting techniques used to derive the
S\'ersic profile parameters appearing in Equation~(\ref{intro}); in
Sections~\ref{CorrelationsR} and \ref{CorrelationsV} we look for
correlations between these parameters; in Section~\ref{Another
relation?} we show that besides the uniqueness of the specific entropy
of galaxies, another relation is also observed. This question is
revisited in the context of the cosmological scenario of hierarchical
merging galaxy formation in Section~\ref{constant?}. We argue that the
observed variations of the specific entropy of galaxies are correlated
with their total luminosity (or mass). We then show in
Section~\ref{entroline}, how this correlation helps to understand the
tilt of the Entropic Line defined in LGM and therefore to further
refine the profile-shape distance indicator of galaxies based on the
shape parameter.  We discuss our results in the last section.

\section{The data and fitting methods}

Our goal is to fit the surface brightness of elliptical galaxies
with two or three parameters, to search for correlations between these
parameters and to look for underlying physical properties.

\subsection{The data} \label{data}

\subsubsection{Real galaxies}

We have used data on galaxies belonging to three clusters: Coma, Abell 85 and
Abell 496. These galaxies were selected: 1) visually as having an elliptical
shape on our CCD images, and 2) spectroscopically as having a redshift within
the corresponding cluster range. The photometric data are described in Lobo et
al. (1997), Slezak et al. (1998) and Slezak et al. (1999), and the spectroscopic
data in Biviano et al. (1995), Durret et al. (1998) and Durret et al. (1999) for
these three clusters respectively. We have determined the growth curve of each
galaxy using the \textsc{ellipse} task of \textsc{iraf}\footnote{\textsc{iraf}
is the Image Analysis and Reduction Facility made available to the astronomical
community by the National Optical Astronomy Observatories, which are operated by
the Association of Universities for Research in Astronomy (AURA), Inc., under
contract with the U.S. National Science Foundation.}. The growth curves were
determined with and without background subtraction.

The first two of these clusters have already been analyzed for the same purpose
in a previous paper (LGM), but with circular apertures. We present here a new
analysis of these two clusters, together with a third cluster (Abell 496), based
on elliptical fits adapted to the geometry of each galaxy.

\subsubsection{Simulated (\textit{virtual}) galaxies}

We have used the merger remnants described in Capelato et al. (1995,
1997). There are three merger generations: (1) the end-products of
merging King spheres (with varying impact parameter, relative energy
and angular momentum), (2) mergers between first-generation mergers,
and (3) mergers between second-generation and between first and
second-generation mergers.

\subsection{The fitting method}

\subsubsection{Real galaxies}

For each galaxy in our cluster sample, we have obtained the growth
curve (integrated luminosity within elliptical regions of area
$\epsilon \equiv \pi A \times B$, where $A$ and $B$ are the semi-major
and semi-minor axes).

The background contribution was determined individually for each galaxy by
fitting the last four points of the growth curve as a function of the surface by
a straight line. We checked the robustness of this result by also fitting the
last 5, 3 and 2 points of the growth curve.

After determining the background contribution, we have subtracted it
from the growth curve. Then we determined the total luminosity,
$L_{\rm tot}$, using the last points of the growth curve, the
half-luminosity (or effective) radius, $R_{\rm eff}$, and the radius
containing 99\% of the total luminosity, $R_{99}$.

We have then fit the growth curves (corrected for the sky) using the
integrated form of the S\'ersic law:
\begin{equation}
L(R)=\frac{2 \pi a^2}{\nu} \; \Sigma _{0} \; \gamma\! \left(
\frac{2}{\nu }, \left( \frac{R}{a} \right)^\nu \right)
\label{Lsersic}
\end{equation}
where the value of $R$ is not a radius but an equivalent radius, $R= \sqrt{A
B}$, and $\gamma(c, x)$ is the standard incomplete Gamma function. The
luminosity growth curve fits were done with a standard least square minimization
method, using the `\textsc{minuit}' programme from the CERN software library.

In order to avoid effects due to the seeing, we have used only data points from
2.0 arcsec outwards (the seeing was FWHM $\approx 0.9$ arcsec for Coma and 1.2
arcsec for A85 and A496). The fits were done using data points up to $R_{99}$ so
that for all galaxies the same amount of light was used for the fits. The
results of these fits are given in Tables~\ref{comafit}, \ref{a85fit}, and
\ref{a496fit}.

\subsubsection{Simulated galaxies}

For simulated galaxies we have computed a mean radial mass growth
curve (that is, integrated mass instead of integrated luminosity), as
follows.  Each galaxy was randomly projected in a plane and a growth
curve was computed for each projection by simply counting the
particles inside iso-density ellipses. These projected growth curves
(500 for each galaxy) were then used to compute a mean one. Having
determined a growth curve for each simulated galaxy, we have proceeded
in the same way as for \textit{real} galaxies, fitting it to the
integrated S\'ersic profile.

The effective radius $R_{\rm eff}$ (i.e. the projected radius containing half of the
total light) is given by LGM:
\begin{eqnarray}
R_{\rm eff} &=&a\, R^*_{\rm eff} \, ;\nonumber  \\
\ln(R^*_{\rm eff}) &=& \frac{0.70348 -0.99625 \ln \nu}{\nu} -0.18722 \, .
\label{reff}
\end{eqnarray}

\begin{figure}[htb]
\includegraphics[width=7.5cm]{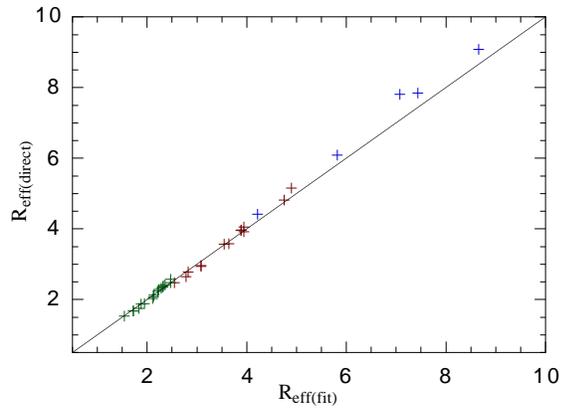}
\caption{Effective radius directly measured on the simulation (R$_{\rm eff
(direct)}$) versus effective radius calculated from the S\'ersic fit (R$_{\rm
eff (fit)}$). The straight line indicates the equivalence between both effective 
radii.}
\label{reffhugo}
\end{figure}

We notice that the simulated galaxy profiles we obtained were
extremely close to S\'ersic profiles. This agrees with the results
obtained by Capelato et al. (1995) using a different fitting
technique. In Fig. \ref{reffhugo} we compare the effective radius of
simulated galaxies as estimated from relation (\ref{reff}) with that
directly measured on the simulations. As it can be seen there is a
good agreement between these two quantities.

\section{Relations and correlations}

The last columns of Tables \ref{comafit}, \ref{a85fit}, and
\ref{a496fit} give the values of the specific entropy, $s_{0}$, for the
galaxies of our cluster sample. Notice that we cannot compare directly
the values of $s_{0}$ since they depend on $a$, which is distance
dependent (we use the apparent $a$ given in arcsec).
Neither can we compare the values of $s_{0}$ for a real
cluster and for the simulation, since they have different units for $a$
and $\Sigma_{0}$. However, we can compare different generations of
mergers, as well as the relative dispersion around the mean value of $s_{0}$
for all real and simulated galaxies.
 
Table \ref{stat_gal} gives the mean values and dispersions of the
specific entropy for the whole sample of galaxies. As expected, the mean
specific entropy varies from cluster to cluster, reflecting the
different distance of each cluster (cf. LGM).  For each real cluster,
the dispersion is about 10\% around the mean value.
 
For \textit{virtual} galaxies, the specific entropy seems to increase
with the hierarchy, but with a much smaller dispersion, around 5\% of
their mean values within each generation. This small dispersion is
reminiscent of the results discussed by Capelato et al. (1997), which
show that the scatter of the Fundamental Plane defined by
\textit{virtual} galaxies is smaller than for the observed ones by a
factor of about 2. The increasing of $s_0$ with the hierarchy of the
merger will be addressed in Section~\ref{constant?}.

\begin{table}[htbp]
\centering
\caption{Mean specific entropy statistics.}
\tabcolsep=0.6\tabcolsep
\begin{tabular}{rccc|cccc}
    \hline
    & \multicolumn{3}{c}{Real galaxies}&\multicolumn{4}{c}{Virtual galaxies}\\
    & Coma   &   A85  &  A496  & 1$^{\rm st}$ gen & 2$^{\rm nd}$ gen & 3$^{\rm rd}$ gen  & All \\
    \hline
Npts   & 69   &  30  &  34  &  17  &  13  &  5   &  35 \\
Mean   & -7.7 & -8.9 & -8.7 & 3.59 & 4.31 & 4.96 & 4.05 \\
Median & -7.8 & -9.2 & -8.9 & 3.61 & 4.34 & 4.91 & 4.07 \\
$\sigma_{s_{0}}$ & 0.7 & 0.8 & 0.8 & 0.09 & 0.15 & 0.29 & 0.53 \\
    \hline
\end{tabular}
\label{stat_gal}
\end{table}

It is the relatively small scatter of the specific entropy of galaxies
around their mean values which justifies the results discussed in LGM,
leading to the definition of a mean specific entropy plane, defined through
Equation (\ref{intro}). We will call it the Entropic Plane.
However, as we will see in the following, there is another relation
linking the observed quantities of galaxies.

\subsection{Correlations with \textit{real} galaxies} \label{CorrelationsR}

\begin{figure}[htb]
\includegraphics[width=7cm]{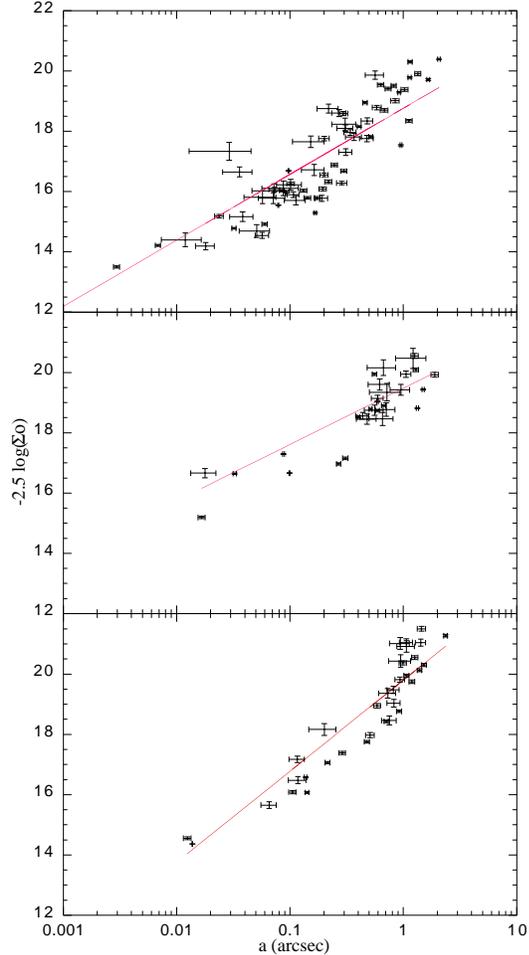}
\caption[]{The correlation between $a$ and $\Sigma_{0}$ for galaxies
belonging to the following clusters: (top) Coma; (middle) Abell 85;
(bottom) Abell 496. The lines are those discussed in Section 
\ref{Another relation?} and Appendix~II.}
\label{asigma}
\end{figure}

\begin{figure}[htb]
\includegraphics[width=7cm]{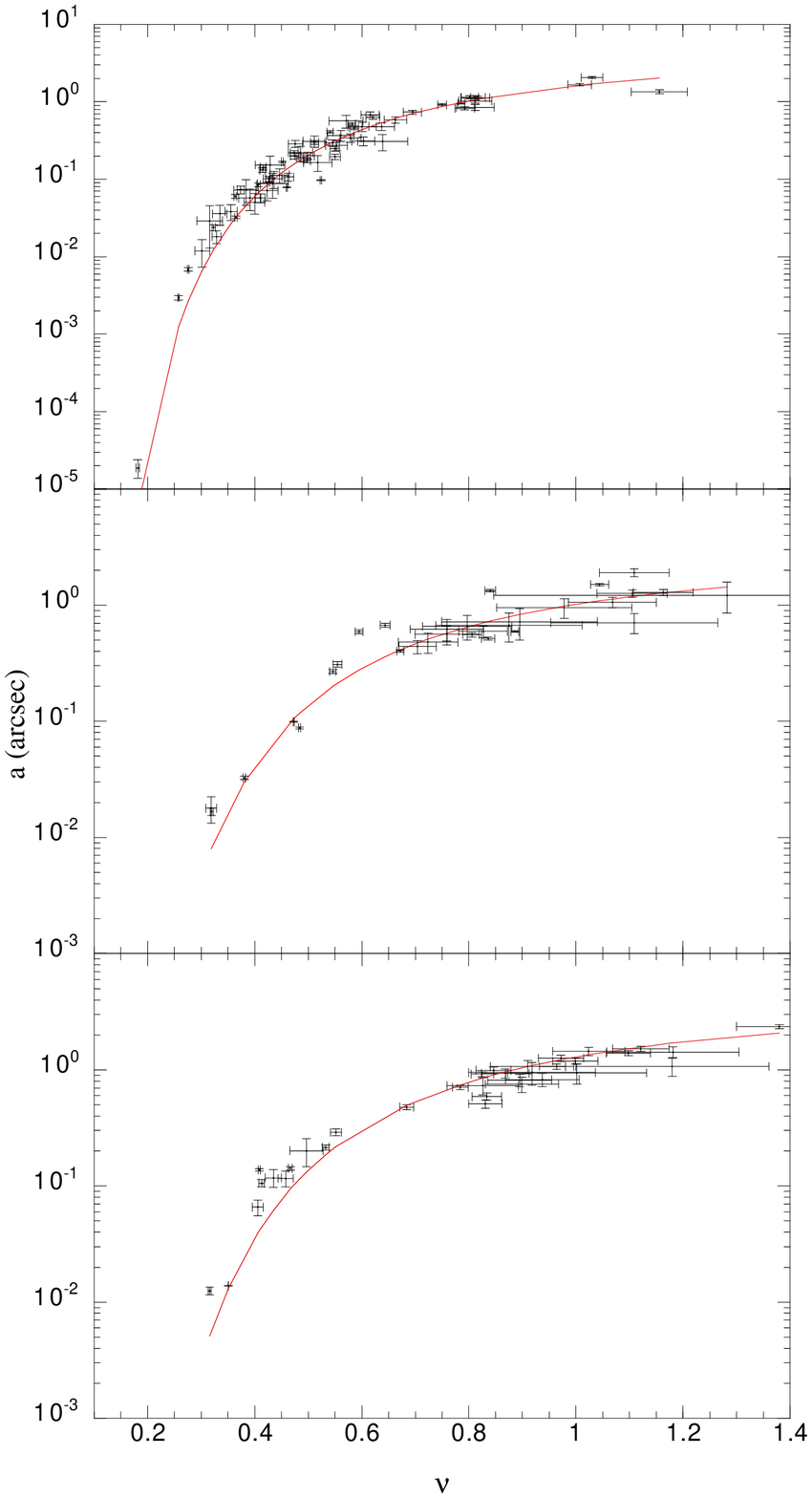}
\caption[]{Same as Fig.~ \ref{asigma} but for $\nu$ and $a$. Top panel: Coma;
middle: Abell 85; bottom: Abell 496.}
\label{anu}
\end{figure}

\begin{figure}[htb]
\includegraphics[width=7cm]{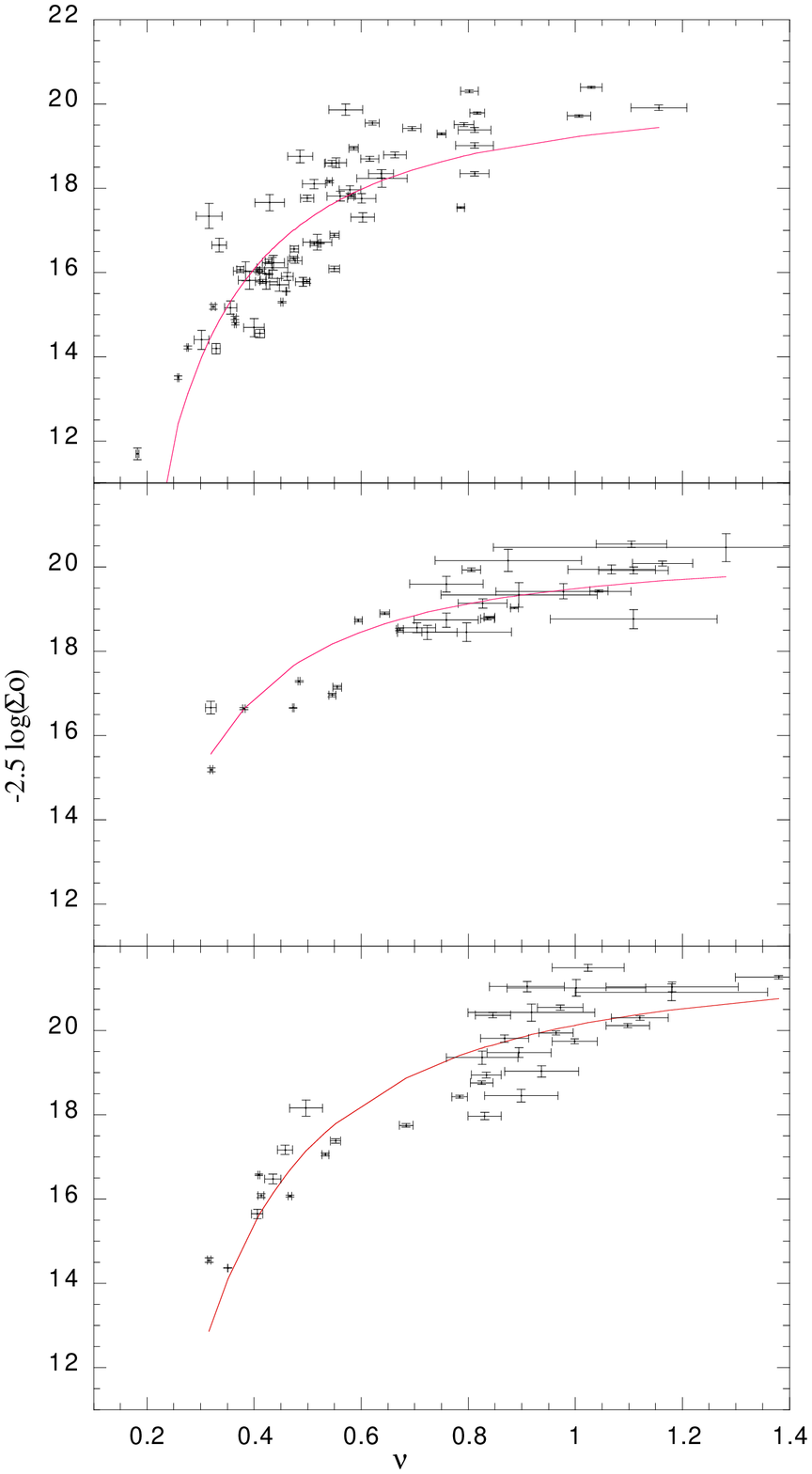}
\caption[]{Same as Fig.~ \ref{asigma} but for $\nu$ and $\Sigma_{0}$. Top panel: Coma;
middle: Abell 85; bottom: Abell 496.}
\label{nusigma}
\end{figure}

The correlations of the S\'ersic profile parameters taken two by two are
displayed in Figures~\ref{asigma}, \ref{anu} and \ref{nusigma}. Notice
that we used the ``astronomical like'' quantity $-2.5 \log \Sigma_0$
instead of $\Sigma_0$.

\begin{figure}[htb]
\includegraphics[width=7cm]{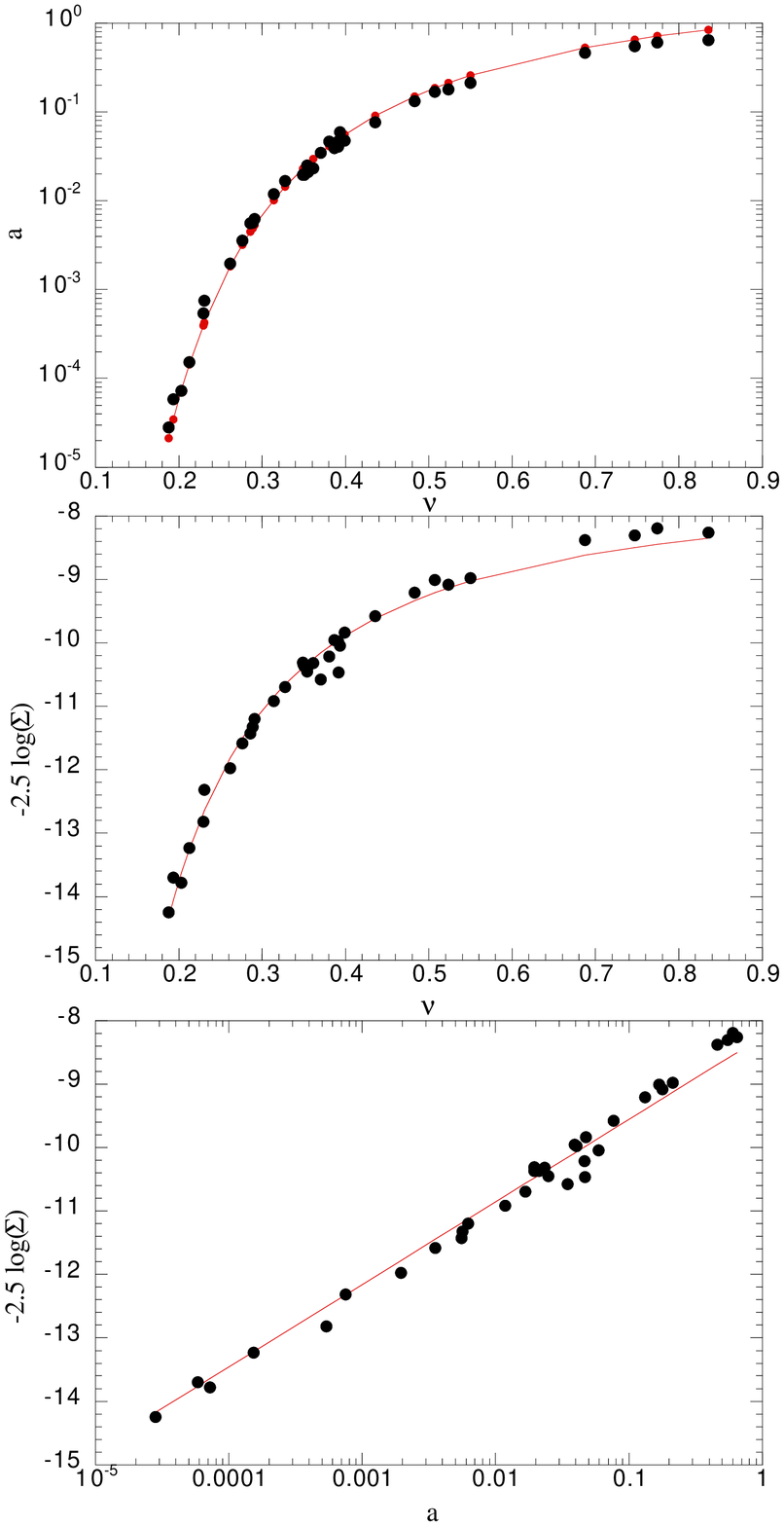}
\caption[]{The correlations between the S\'ersic profile parameters for
the simulated galaxies. The lines have the same meaning as
for Figs.~ \ref{asigma}, \ref{anu} and \ref{nusigma}.}
\label{correlhugo}
\end{figure}

These correlations all have the same general aspect as those presented
in LGM: the three parameters appear well correlated two by two. This
strongly suggests that galaxies are not distributed randomly around their
mean Entropic Plane but, instead, are distributed along a
thin curve in this plane.

\subsection{Correlations with \textit{virtual} galaxies}\label{CorrelationsV}

A similar analysis applied to \textit{virtual} galaxies is
shown in Figure~\ref{correlhugo}. The correlations of the S\'ersic
parameters taken two by two are similar to those found for
\textit{real} galaxies. However, the scatter for \textit{virtual}
galaxies is much smaller.

\subsection{Another relation?}\label{Another relation?}

The results discussed above suggest that we may go one step further:
since the projections of the galaxies belonging to the \textit{Entropic
Plane} are simultaneously three thin curves, the galaxies must in fact
be located on a thin curve in the Entropic Plane.

In order to check this hypothesis, we will look at the
Entropic Plane edge-on and face-on. This requires a rotation
to a new coordinate system [$\xi, \eta, \zeta$], defined by:
\begin{eqnarray}
\xi &=& [\ln (a)-5\ln (\Sigma_{0})]/\sqrt{26} \nonumber \\
\eta &=& [-5\ln (a)-\ln (\Sigma_{0})+13 F(\nu)]/\sqrt{195} \nonumber \\
\zeta &=& [5\ln (a)+\ln (\Sigma_{0})+2 F(\nu)]/\sqrt{30} \, .
\label{rotation}
\end{eqnarray}
%

\begin{figure}[htb]
\includegraphics[width=8.25cm]{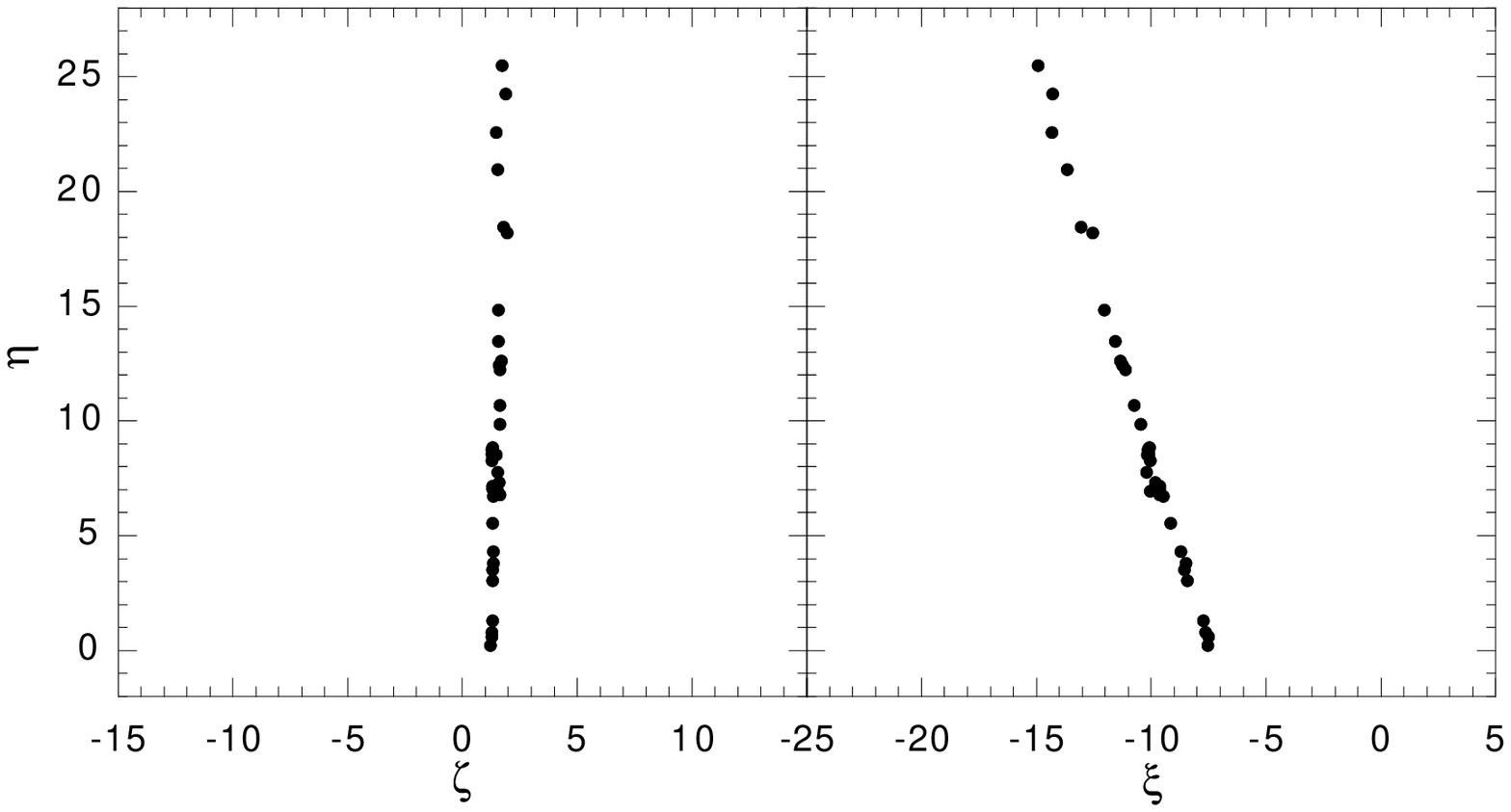}
\caption[]{\textit{Virtual} galaxies. Right panel: Entropic Plane seen
face-on. Left panel: Entropic Plane seen edge-on.
Notice how $\zeta$, which is equivalent 
to $s/\sqrt{7.5}$, has a much smaller variation vis-\`a-vis $\eta$ and $\xi$.}
\label{entropievirtuelle}
\end{figure}

We apply this rotation to each galaxy and show the result in
Fig. \ref{entropievirtuelle} for \textit{virtual} galaxies.
This figure suggests several comments:
\begin{itemize}
\item The value of the specific entropy depends only on $\zeta$, $s=\sqrt{7.5}\,
\zeta$. Indeed, as a first approximation, \textit{virtual} galaxies all have the
same specific entropy, its numerical value depending
on the choice of units. Deviations from this
constant value will be addressed in the next section.

\item The galaxies are all effectively located on a curve (in fact very nearly a
straight line): $$ L(\xi, \eta) \equiv L(\Sigma_{0}, a, \nu) = 0.$$

\end{itemize}

\begin{figure}[htb]
	\includegraphics[width=8.5cm]{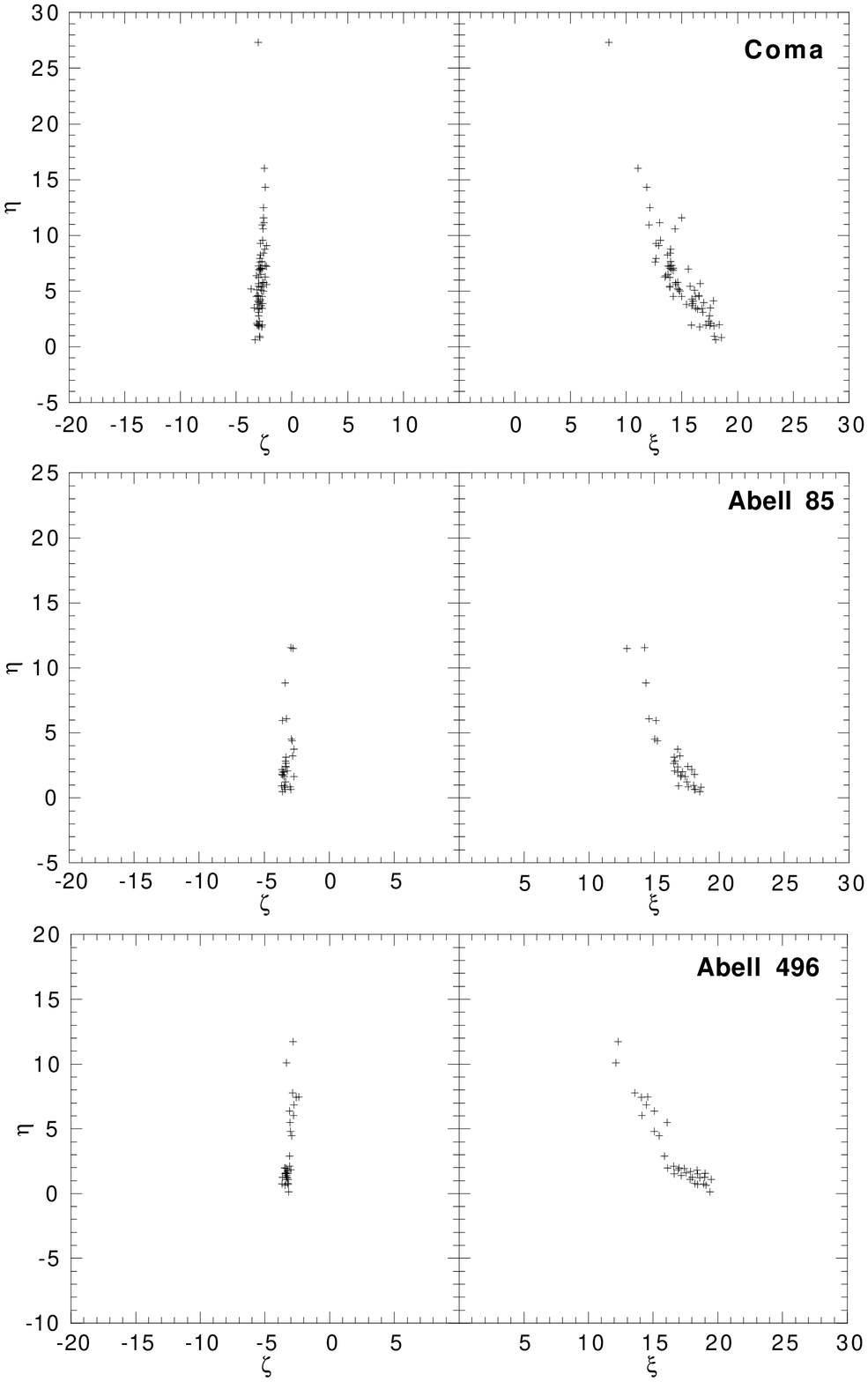}
\caption[]{Right panel: Entropic Plane seen
face-on. Left and bottom panels: Entropic Plane seen
edge-on. Top: Coma, middle: Abell 85, bottom: Abell 496.}
	\label{entropietoutes}
\end{figure}

Applying the same rotation to the \textit{real} galaxies data gives
quite similar results as for the \textit{virtual} ones, as 
seen in Fig. \ref{entropietoutes}. However:
\begin{itemize}
\item  The scatter is much larger;

\item  It is not clear whether the relation $L(\xi, \eta)=0$ found for
\textit{virtual} galaxies can be approximated by a straight line as before.

\end{itemize}

The curves displayed in Figs.~\ref{asigma}--\ref{correlhugo} were
obtained by assuming that the relation $L(\xi, \eta)=0$ may be
approximated by a straight line. Their derivations are given in
Apppendix II.

At present we do not know of any physical explanation for this
relation. In order to have a curve in a three-dimensional space, we
need two surfaces to intersect. If one of them is the Entropic Plane,
then the other one must be derived from an independent relation, for
instance, a scaling law relating the gravitational potential energy to
the total mass, or the depth of the potential well to the mean
potential (as also suggested by Lima Neto et al. 1999b). Such an
intersection of surfaces is actually a geometrical interpretation of
the scaling relations that govern the physics of a given system. This
kind of problem is similar, for example, to the one encountered when
dealing with the origin of the Tully-Fisher relation (e.g. Mo et
al. 1998) or of the luminosity--temperature relation for X-ray
clusters of galaxies (e.g.  Markevitch 1998).

\section{Is the specific entropy really unique?}\label{constant?}

\subsection{The merging scenario}\label{scenario}

As explained above, the \textit{virtual} elliptical galaxies we are
using come from the merging of successive generations. In such merging
processes the energies and masses of the progenitors will be
redistributed in order to generate another elliptical. In this section
we analyze the effect of merging on the value of the specific entropy.

When viewing a movie displaying the simulation of a galaxy merger as
compared to one displaying a cold self-gravitational collisionless
collapse, one is struck by the much more violent matter motions
(which are actually the engine for the violent relaxation process)
occurring in the collapse simulation. Furthermore, what one observes
in such movies is, in a sense, quite similar to an observation of the
real universe, that is, a \textit{macroscopic} observation. This is in
contrast with the microscopic description one may obtain from the
knowledge of detailed evolution of the phase space positions of each
particle provided by the simulation. In other words, the relevant
description of the system is done with a coarse-grained distribution
function.

In any case, the mixing in phase-space that occurs during violent
relaxation is responsible for the increase of the
\textit{coarse-grained} entropy of a dynamical system (Tremaine et
al. 1986, Merritt 1999). However, it is not clear how much the entropy
or, perhaps more important, the specific entropy, increases during the
violent relaxation phase. Does the entropy increase depend on the
amplitude of the time-varying potential induced by the violent matter
motions occurring during this phase? It could be that the increase of
entropy is small compared to the total mass of the system (note that
defining the entropy as in Eq.~(\ref{eq:boltzmann}), $S$ has the unit
of a mass). In this case, the specific entropy increment could be
insignificant when compared with the change of other quantities like,
e.g., the total gravitational energy or the total mass of the system
(as indeed will be shown below).

\subsection{Shift of the specific entropy}

\begin{figure}[htb]
	\includegraphics[width=8cm]{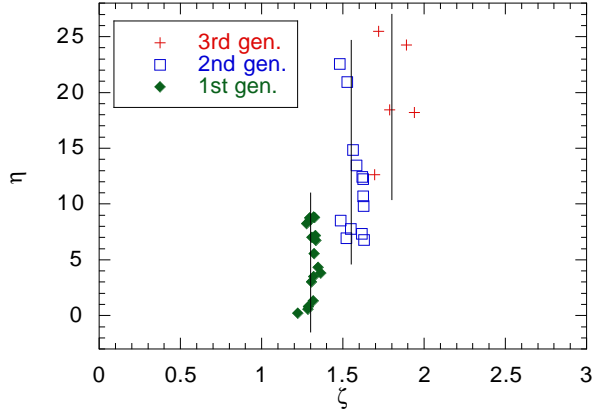}
\caption[]{Enlargement of the $\zeta$--$\eta$ view of the Entropic Plane for 
\textit{virtual} galaxies. Each merger generation is represented by a different 
symbol. The vertical lines show the mean values of $\zeta$ for each generation.}
	\label{gene}
\end{figure}

A careful inspection of the specific entropy (which is given by
$\sqrt{7.5}\,\zeta$) both for \textit{virtual} and \textit{real}
galaxies indeed shows that it does vary, although slightly.  We have
zoomed the left panel of Fig. \ref{entropievirtuelle} and show the
result in Fig. \ref{gene}: an overall increase of the specific entropy
is observed ($\zeta$ is not exactly a constant). Three vertical lines
have been drawn, corresponding to the mean entropy of the three
successive generations of galaxies.  The specific entropy is actually
different for each generation of galaxies, and seems to increase by
quanta of specific entropy from one generation to the next, although
the jump of specific entropy is quite small (between 10 and 20~\%, see
also Table~\ref{stat_gal}).

\begin{figure}[htb]
	\includegraphics[width=8cm]{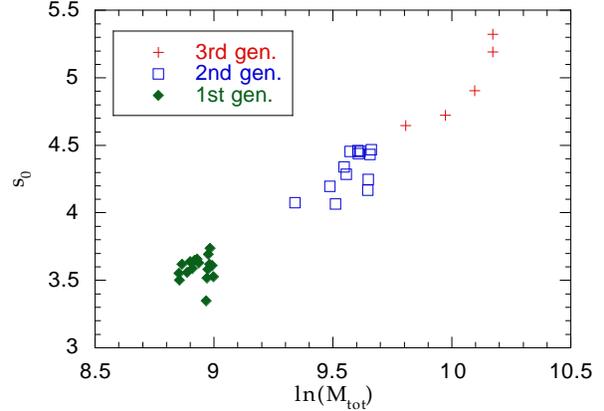}
\caption[]{\textit{Virtual} galaxies. Specific entropy versus total mass. 
Symbols are the same as in Fig.~\ref{gene}.}
	\label{svirmass}
\end{figure}

One obvious difference between galaxies of different generations is the increase 
of their total mass. In fact, even in a given generation there are galaxies with 
different masses because of slightly different initial orbital parameters of 
the progenitors. Therefore, 
we have plotted the specific entropy as a function of the total light
(mass), i.e., the integrated luminosity (mass) given by relation
(\ref{Lsersic}) extrapolated to infinity:
\begin{equation}
    L_{\rm tot} = \frac{2\pi a^2}{\nu}\;\Sigma_{0}\; \Gamma(\frac{2}{\nu})
    \label{ltot}
\end{equation}
The corresponding Fig. \ref{svirmass} shows that the total mass
allows to discriminate clearly the entropy of the three generations of
galaxies. Notice that it is not the mass by itself which is really
responsible for the shift of the specific entropy, but rather the merging
process.

\begin{figure}[htb]
	\includegraphics[width=8cm]{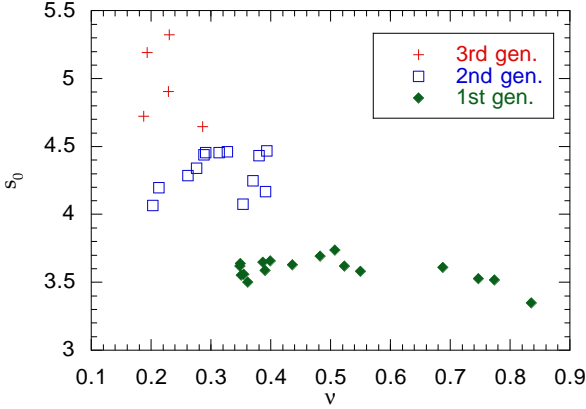}
\caption[]{\textit{Virtual} galaxies. Specific entropy versus $\nu$.
Generation symbols are the same as in Fig.~\ref{gene}.}
	\label{svirnu}
\end{figure}

Other parameters do not allow such a clear discrimination between
generations. For instance we have plotted the specific entropy as a
function of the $\nu$ parameter for \textit{virtual} galaxies
(Fig. \ref{svirnu}). For a given value of $\nu$, several values of the
specific entropy are possible, implying that the parameter $\nu$ is not a
good discriminant between generations.

\begin{figure}[htb]
	\includegraphics*[width=6.2cm]{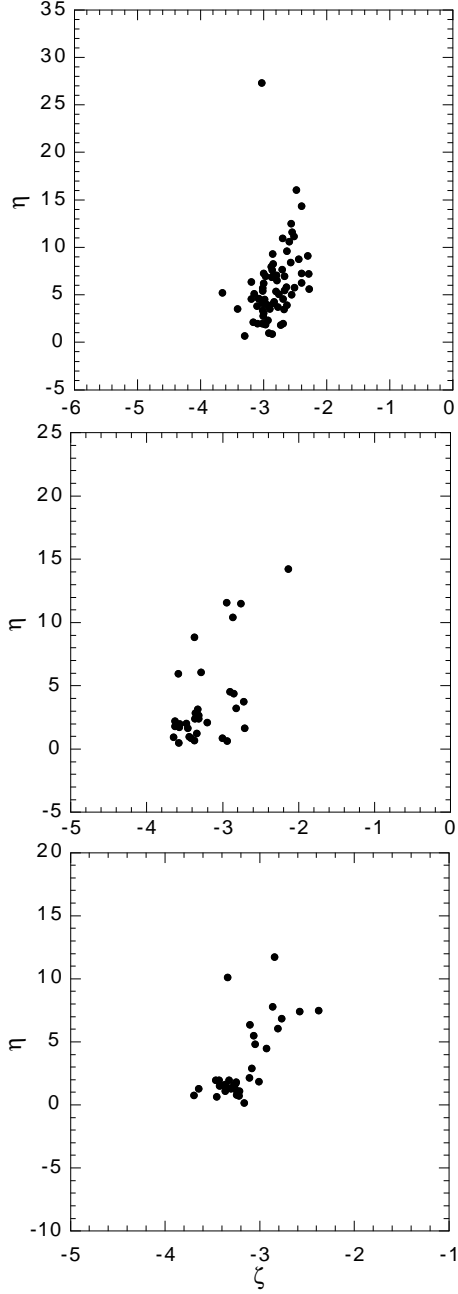}
\caption[]{Rotated Entropic Plane for \textit{real} galaxies.
Upper panel: Coma Cluster.
Middle panel: Abell~85. Lower panel: Abell~496. Notice that the $\zeta$ axis is 
enlarged. This figure is comparable  to 
Fig.~\ref{gene} (keeping in mind the different units).}
	\label{gene2}
\end{figure}

\begin{figure}[htb]
	\includegraphics[width=8cm]{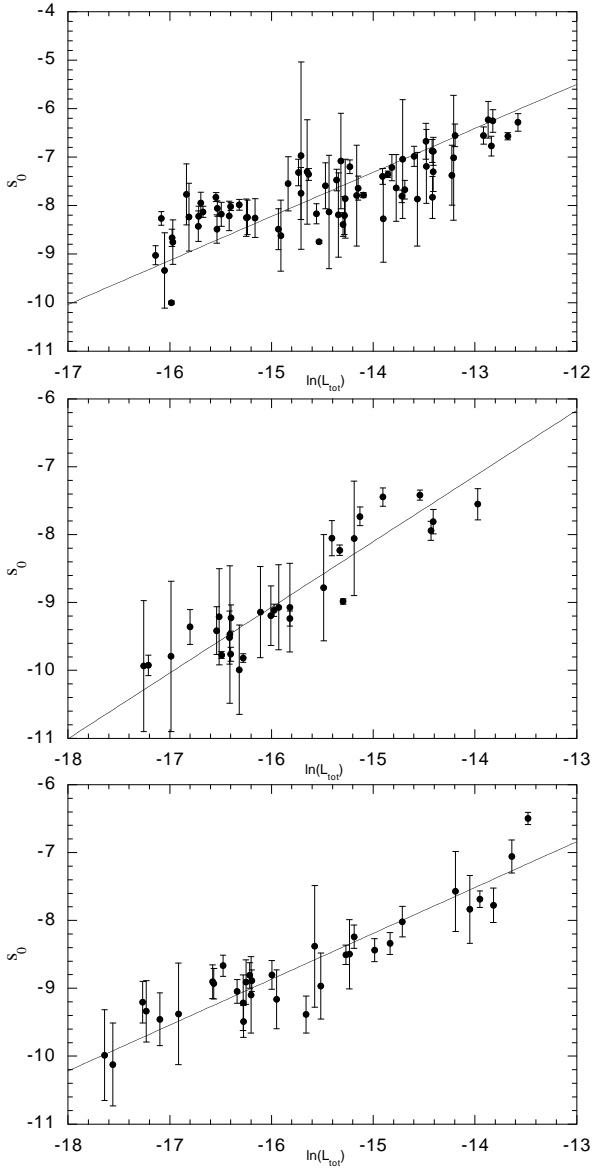}
\caption[]{Correlation between the specific entropy and the total luminosity 
(mass) for {\textit real} galaxies. Top panel: Coma Cluster.
Middle panel: Abell~85. Bottom panel: Abell~496.}
	\label{sreall}
\end{figure}

\begin{figure}[htb]
	\includegraphics[width=8.4cm]{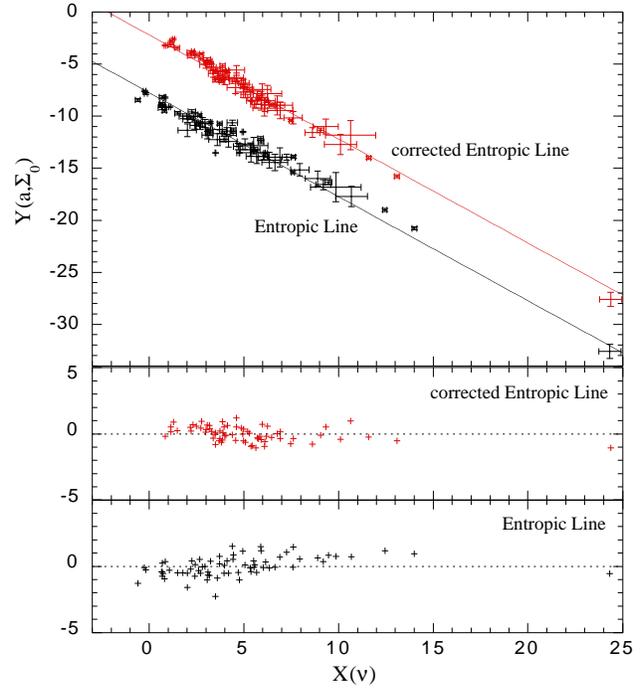}
\caption[]{Coma Cluster. Top panel: \EL\ and \CEL\ (arbitrarily separated
by a constant value to facilitate the comparison). The straight lines have a 
slope of $-1$, as predicted by Eqs.~(\ref{ligne}) and (\ref{lignecor}).
Bottom panel: residuals.}
\label{resentropiecoma}
\end{figure}

\begin{figure}[htb]
	\includegraphics[width=8.4cm]{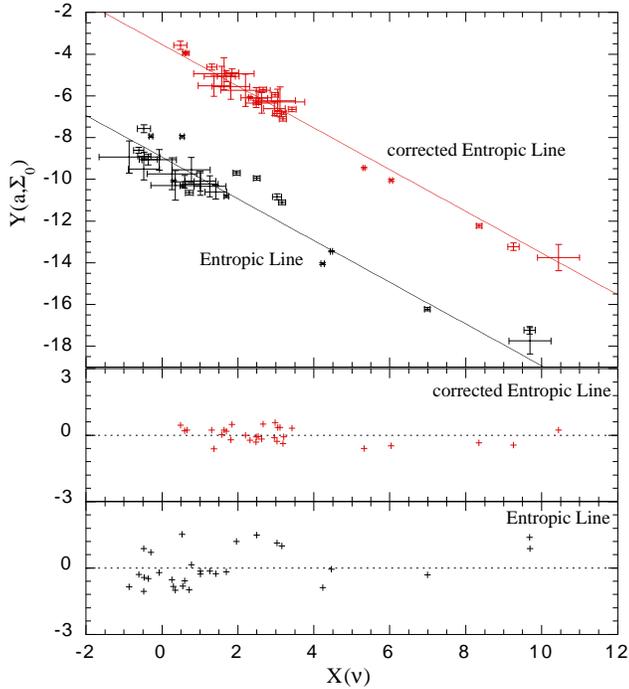}
\caption[]{Same as in Fig.~\ref{resentropiecoma} for Abell~85.}
	\label{resentropieA85}
\end{figure}

\begin{figure}[htb]
	\includegraphics[width=8.4cm]{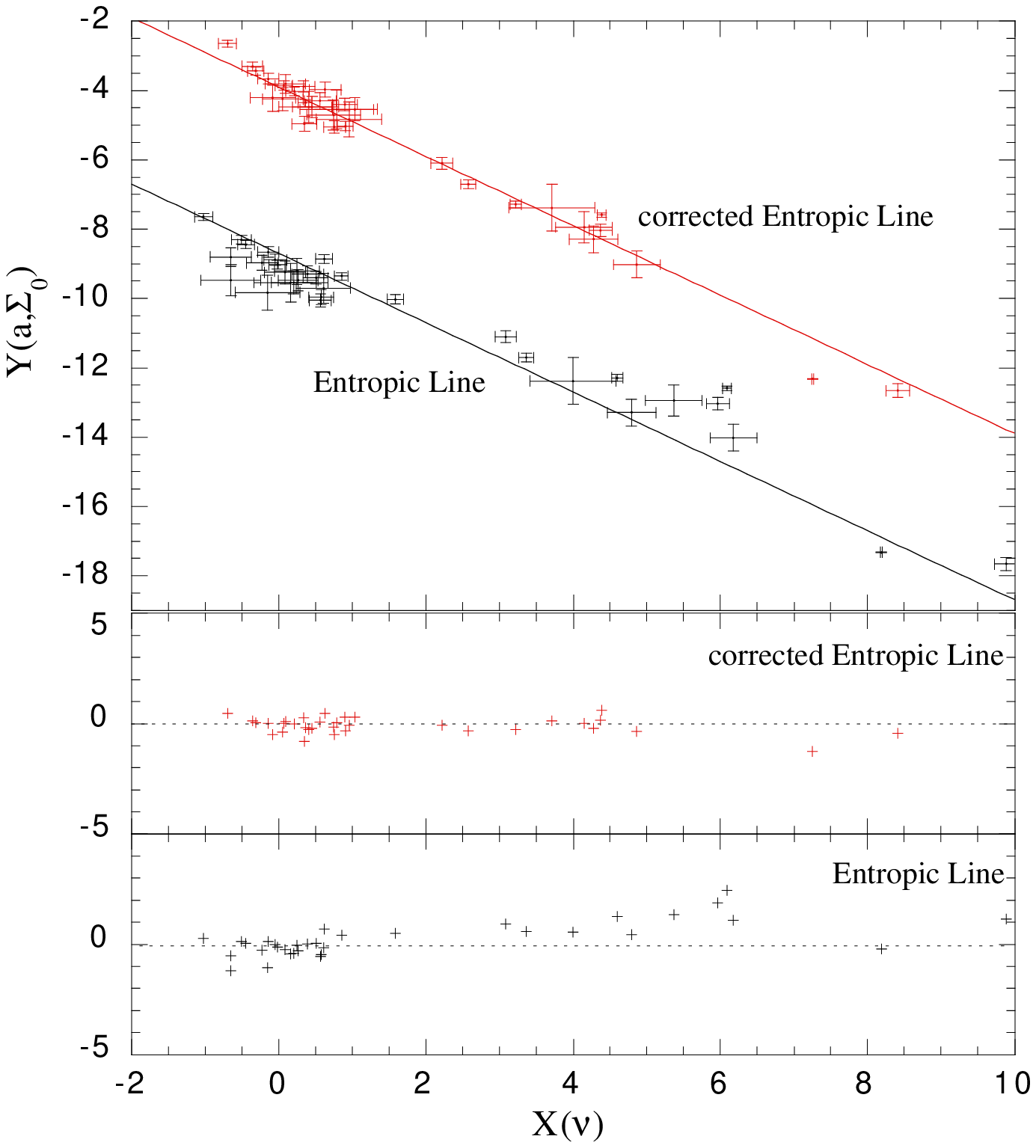}
\caption[]{Same as in Fig.~\ref{resentropiecoma} for Abell~496.}
	\label{resentropieA496}
\end{figure}

In Fig. \ref{gene2} we show a plot similar to Fig.~\ref{gene}, but now
using the \textit{real} galaxies data; the same overall trend of an
increase of the specific entropy ($\zeta$) with $\eta$ seems to occur
with \textit{real} galaxies. The striking similarities of these two
figures further reinforces the hypothesis that elliptical galaxies
have been formed by mergers.

Although we do not know to which generation each \textit{real} galaxy
belongs, we may push the analogy between \textit{real} and
\textit{virtual} galaxies a little further by searching for
correlations between the entropy and luminosity. The corresponding
plots are shown in Fig. \ref{sreall}. Again, there is a correlation
between $s_{0}$ and $L_{\rm tot}$ for \textit{real} galaxies, although
with lower signal-to-noise ratio. We now investigate the implications
of such a correlation in the context of the Entropic Line defined by
LGM.

\section{The Entropic Line revisited}\label{entroline}

LGM have proposed to rewrite Equation (\ref{intro}) as:
\begin{eqnarray}
Y(\Sigma_{0}, a)&=& 0.5\ln(\Sigma_0) + 2.5 \ln(a) \, ;\nonumber \\
X(\nu) &=& F(\nu) \, ;\nonumber \\
Y + X &=& s_{0} \, .
\label{ligne}
\end{eqnarray}

Equation (\ref{ligne}), relating $Y$ as a function of $X$, is the
equation of a straight line with a slope $-1$, called the Entropic
Line in LGM, which is, in fact, the Entropic Plane seen edge-on; in
this equation $s_0$ is the mean value of the specific entropy of
galaxies in a given cluster (as in Table~\ref{stat_gal}). Remember
that $s_{0}$ is a distance dependent quantity and thus that values for
different clusters are not directly comparable.

In their paper, LGM compared the observed data to the predictions of
Eq.~(\ref{ligne}), showing that there is a small difference between
the corresponding slopes (see Table \ref{stats}, col.~3). The fact
that the Entropic Line obtained with the observed data had a slope
different from $-1$ was referred to as the tilt of the Entropic
Line. However, as we have seen in the last Section, the specific
entropy varies with total mass (or light) and, as we will see below,
this may be at the origin of the tilt.

\begin{table}[htb]
\caption[]{Statistical analysis: Coma, Abell 85 and Abell 496. Col. 2:
mean specific entropy (note that $s_{0}$ is not directly comparable for
different clusters, see text); col. 3: slope of the correlation of
$Y(\Sigma_{0}, a)$ with $X(\nu)$; col. 4: slope of the correlation of
$Y(\Sigma_{0}, a)$ with $X_{cor}(\nu)$; col. 5: standard deviation of the data
residuals relative to the \EL; col. 6: standard deviation of the data residuals
relative to the \CEL.}
\tabcolsep=0.6\tabcolsep
\begin{tabular}{lccccc}
    \hline
  Cluster & $s_0$ & slope$_1$ & slope$_2$ & $\sigma_1$(residual) & 
  $\sigma_2$(residual) \\
    \hline
   Coma      & -7.7 & -0.93 & -1.06 & 0.74 & 0.51 \\
   Abell 85  & -8.9 & -0.88 & -1.05 & 0.82 & 0.34 \\
   Abell 496 & -8.7 & -0.81 & -1.05 & 0.78 & 0.37 \\
    \hline
\end{tabular}
\label{stats}
\end{table}


We assume the specific entropy $s_{0}$ to be a function of the total luminosity,
$L_{\rm tot}$, as given by Eq.~\ref{ltot}:
$$s_{0} = \alpha \ln(L_{\rm tot}) + s_{0,0}$$
the slope $\alpha$ being obtained through an ordinary least square mean 
fitting of the data (shown as lines in Fig. \ref{sreall}). The
intercept $s_{0,0}$ may be understood as a corrected specific entropy, that is,
taking into account the correlation with $L_{\rm tot}$.

A ``corrected'' equation can then be written in place of equation
(\ref{ligne}) as:
\begin{eqnarray}
X_{cor}(\nu) &=& F(\nu)-\alpha \ln(L_{\rm tot}) \nonumber \\
Y+X_{cor}&=&s_{0,0}
\label{lignecor}
\end{eqnarray}

In Figs. \ref{resentropiecoma}, \ref{resentropieA85} and \ref{resentropieA496}
we display both the original LGM Entropic Line and the \textit{corrected}
Entropic Line, together with their corresponding residuals, for the clusters of
our sample. Notice that the slope of the \textit{corrected} Entropic Line is
close to $-1$, indicating that the tilt between the data and the predictions
based on an unique specific entropy for galaxies, has diminished (See
Table~\ref{stats}, col.~4). Moreover, the dispersion around the
\textit{corrected} Entropic Line is improved compared to the dispersion around
the uncorrected one (compare cols.~5 and 6 in Table~\ref{stats}).

\section{Discussion and conclusions}

We have shown in this paper that the observational data for galaxies belonging
to clusters confirm the hypothesis that their specific entropy is, to first
order, unique. As a consequence, the galaxies tend to stay in a thin plane (the
Entropic Plane) in the space of the S\'ersic light profile parameters. Moreover,
we have also shown that the slight observed variations of the specific entropy
of galaxies are correlated to their total luminosities. Henceforth, by taking
this correlation into account, we were able to apply a correction which resulted
in decreasing the scatter of the Entropic Plane.

We have also shown that, besides the Entropic Plane, another relation
must exist between the S\'ersic profile parameters of galaxies. The
intersection of this relationship with the Entropic Plane defines a
curve in the 3D S\'ersic parameter space which appears very well
defined in our data. The physical origin of this new relation is
unknown, although it may be related to the gravitational energy of
galaxies, as suggested by Lima Neto et al. (1999b).

The existence of this curve explains the tight correlations between
the \Se\ parameters, as noted in LGM. It constitutes the theoretical
background for the photometrical distance indicator proposed by
various authors (Young \& Currie 1994, 1995; Binggeli \& Jerjen 1998).

Our study is solely based on quantities extracted from photometric data. The
dynamics are in fact \textit{hidden} in the shape parameter of the fitting
S\'ersic profiles. Usually, both photometric and spectroscopic studies are
performed on galaxies, leading for instance to the ``Fundamental Plane". It is
interesting to notice that some authors have shown that globular clusters
(Bellazzini 1998), galaxies (Guzm\'an et al. 1993) or even galaxy clusters (Fujita
\& Takahara 1999) are located on a line in that plane instead of populating the
whole area \textit{a priori} permitted by the natural range of variation of the
parameters. The question of how to derive the ``Fundamental Plane'' (photometry
plus spectroscopy) from the Entropic Plane (photometry alone) will be addressed
in a forthcoming paper.

Elliptical galaxies are well described by a three parameter profile;
since two constraints are acting, only one parameter is free. Because
the total luminosity of galaxies correlates with the shape parameter
(Prugniel \& Simien 1997, Binggeli \& Jerjen 1998), their light
profile should be completely defined by their total luminosity. Shape
and luminosity are probably correlated as a consequence of the
relaxation process; this is consistent with the fact that high
luminosity galaxies have flat profiles while those of dwarfs are
peaked.

We must stress that the calculation of the specific entropy requires a good
choice for the density profile. Indeed, besides an intensity and a length scale,
which are the basic parameters of all the laws used (for instance the \DV\ law),
structural parameters are also necessary in order to account for the shapes,
which reflect the dynamical properties of the galaxies.

In this context, we found that the choice of the \Se\ law is particularly
suitable not only for the sake of simplicity, but also because it is the bulk of
the light which is affected by the variations of the shape factor, as pointed by
Graham \& Colless (1997). One can find in the literature other profiles
depending on a shape factor, for instance the $\beta$-model. However, due to the
asymptotic behaviour of this model, the entropy and mass calculated in this case
are essentially located at large radial distances. Besides, the values of these
two quantities are very sensitive to any computational cut-off, and, moreover,
the observational data in the outskirts of clusters are generally too poor to
guide the calculation. Therefore, the $\beta$-model is not suitable for such
calculations.


We do not claim that the \Se\ profile is the ultimate profile of relaxed
systems, any more than for instance the \DV\ profile. However, while the precise
analytical expression for a profile may not be fundamental, we understand that
taking into account the shape parameters is indeed important. It would then be
interesting to see if ``Universal" profiles -- as for instance the NFW (Navarro
et al. 1995) or the Hernquist (1990) profiles -- also have shape parameters,
allowing the calculation of the specific entropy.

\begin{acknowledgements}
This work was partially supported by
FAPESP, CNPq, PRONEX-246, USP/COFECUB, CNPq/CNRS Bilateral Cooperation
Agreements, and DGCyT grant PB93-0139.
We also acknowledge financial support from the French Programme National de
Cosmologie, CNRS.
\end{acknowledgements}


\section*{Appendix I: Entropy and the S\'ersic profile}

We use the microscopic Boltzmann-Gibbs definition of the specific entropy,
\begin{equation}
    s \equiv S/M =  -\frac{1}{M} \int f(\epsilon) \ln f(\epsilon)
\diff \vec{r}\, \diff \vec{v}
    \label{eq:boltzmann}
\end{equation}
where $M$ is the total mass and we assume that the function $f$ (the
coarse-grained distribution function) depends only on the energy $\epsilon$.
Then, $f(\epsilon)$ can be determined by the Abel inversion of the density
profile (cf. Binney \& Tremaine 1987) and the specific entropy may be computed
(see formulae below).

In order to compute $s$ we have adopted the following hypotheses:

\begin{itemize}
    \item Spherical symmetry;

    \item Isotropy of the velocity distribution;

    \item $M/L(r) = $ constant.
\end{itemize}

From the 2D mass distribution given by the S\'ersic profile (using the
hypothesis of constant $M/L$ ratio), we have derived a semi-analytical
approximation for the 3D mass distribution obtained by deprojecting the
S\'ersic profile (see LGM):
$$\rho (r)=\rho _{0}\left( \frac{r}{a}\right) ^{-p}\exp (-[r/a]^{\nu})\, ;$$
$$ p=1.0-0.6097\nu +0.05563\nu ^{2} \, .$$

From the 3D mass distribution we can compute the distribution function and thus
the specific entropy. The computation can be done numerically but it is
cumbersome. We have therefore found analytical approximations for the specific
entropy (see also LGM):
$$s(a, \nu, \Sigma_0) = \frac{1}{2}\ln \Sigma_{0} + \frac{5}{2}\ln a + F(\nu) 
\, ;$$
$$F(\nu) = 0.2\ln (\nu) - \frac{1.3}{\nu} + 3.9\nu^{-1.3}-2.7 \, .$$
If  $s(a, \nu, \Sigma_0) = s_{0} =$ constant, then the above equations define a 
surface, the Specific ``Entropy Plane'' (in the appropriate variables).

Finally, we give for completeness an analytical approximation for the 
corresponding magnitude for a given set of S\'ersic parameters $(a, \nu, 
\Sigma_0)$:

$$m = -2.5\log L(R\rightarrow\infty) = -2.5\log \Sigma_{0}-5\log a+m^{*} \, ;$$
$$m^{*} = -0.304\nu -1.708\nu ^{-1.44} \, .$$

\section*{Appendix II: Analytical formulae for the correlations}

The projection of the relation $L(\xi, \eta)=0$, introduced in
Section~\ref{Another relation?}, leads to one dimensional curves in the planes
[$\Sigma_{0}, a$], [$\Sigma_{0}, \nu $] and [$a, \nu $]. We give below semi-theoretical
formul\ae\ for these relations:
\begin{enumerate}

\item we assume that $L(\xi, \eta)=0$ may be approximated by a straight line:
\begin{equation}
\eta = A \xi + B \, ,
\label{droite}
\end{equation}
where the constants $A$ and $B$ are obtained through a fitting of the
data;

\item  $\xi$ and $\eta$ depend on $a$, $\nu$ and $\Sigma_{0}$ by relations
(\ref{rotation});

\item we have postulated that the specific entropy is unique
(theoretical aspect) i.e. $\zeta = s_{0}/\sqrt{7.5}$ ($\zeta$ also given by
Eq.~(\ref{rotation})).

\end{enumerate}

Combining (6) and (\ref{droite}) allows us to recover analytical
formulae, which are displayed below. We have superimposed these
calculated curves on each of the corresponding data in
Figs. \ref{asigma}, \ref{anu} and \ref{nusigma}, for the \textit{real}
galaxies and in Fig. \ref{correlhugo} for the simulated ones.

We obtain for the \textit{virtual} galaxies the formula:
\begin{eqnarray}
a &=& \exp[1.771\times (8.256-2.5\times \log \Sigma_{0})]\nonumber \\
-2.5 \log \Sigma_{0} &=& \frac{0.328}{\nu} -
\frac{0.9833}{\nu^{1.3}} - 0.0504 \ln(\nu) -7.5123 \nonumber \\
a &=& \exp[ \frac{0.5804}{\nu} -
\frac{1.741}{\nu^{1.3}} - 0.089 \ln(\nu) + 1.317 ] \nonumber \\
\label{phenoHugo}
\end{eqnarray}

and successively for Coma:
\begin{eqnarray}
a &=& \exp[0.951\times (-18.76-2.5\times \log \Sigma_{0})] \nonumber \\
-2.5 \log \Sigma_{0} &=& \frac{0.6}{\nu} - \frac{1.798}{\nu^{1.3}} 
- 0.0922 \ln(\nu) + 20.42 \nonumber \\
a &=& \exp[ \frac{0.6304}{\nu}
- \frac{1.89}{\nu^{1.3}} - 0.097\ln(\nu) + 1.741] \nonumber \\
\label{phenoComa}
\end{eqnarray}

Abell~85:
\begin{eqnarray}
a &=& \exp[0.811\times (-19.48-2.5\times \log \Sigma_{0})] \nonumber \\
-2.5 \log \Sigma_{0} &=& \frac{0.496}{\nu} -
\frac{1.487}{\nu^{1.3}} - 0.0763 \ln(\nu) + 20.483 \nonumber \\
a &=& \exp[ \frac{0.611}{\nu} - \frac{1.834}{\nu^{1.3}} - 0.094
\ln(\nu) + 1.238] \nonumber \\
\label{phenoA85}
\end{eqnarray}

and Abell~496:
\begin{eqnarray}
a &=& \exp[1.3131\times (-19.80-2.5\times \log \Sigma_{0})] \nonumber \\
-2.5 \log \Sigma_{0} &=& \frac{0.901}{\nu} - 
\frac{2.702}{\nu^{1.3}} - 0.1386 \ln(\nu)+ 21.9 \nonumber \\
a &=& \exp[ \frac{0.686}{\nu} - 
\frac{2.0577}{\nu^{1.3}} - 0.105 \ln(\nu) + 1.6255]\nonumber \\
\label{phenoA496}
\end{eqnarray}

\section*{Appendix III: Growth curve fitting results}

\begin{table*}[htbp]
    \caption[]{Growth curve fitting results for Coma. The Godwin et al. (1983) 
    number is indicated as well as the name from either NGC or IC catalogues.
    $\nu$ and $a$ are the shape
    and scale parameters fit with the \Se\ law. $-2.5\log(\Sigma_0)$ is the
    intensity parameter and the magnitude is in the $V$ band. $R_{\rm eff}$ and
    $\mu_{\rm eff}$ are calculated from the primary parameters. $F(\nu)$ is 
    calculated with Eq.~(\ref{entronu}) and the specific 
    entropy, $s_{0}$, with Eq.~(\ref{intro}).}
    \input{marquez.t1.tex}
    \label{comafit}
\end{table*}

\begin{table*}[htbp]
    \noindent{\bf Table \ref{comafit}.} Continued.\\

    \input{marquez.t2.tex}
\end{table*}

\begin{table*}[htbp]
    \caption[]{Same as Table \ref{comafit} for Abell 85 data. The 
    identification number is that of Durret et al. (1998).}
    \input{marquez.t3.tex}
    \label{a85fit}
\end{table*}

\begin{table*}[htbp]
	\caption[]{Same as Table \ref{comafit} for Abell 496 data. The 
    identification number is that of Durret et al. (1999).}
	\input{marquez.t4.tex}
	\label{a496fit}
\end{table*}

\end{document}

%% file: marquez.t1.tex
\tabcolsep=0.6\tabcolsep
\begin{tabular}{l l c c c c c c c r}
\hline
GMP  & Name     &     $\nu$     &      $\log a$  &  $-2.5 \log \Sigma_0$  &  $L_{\rm Tot}$ & $R_{\rm eff}$ & $\mu_{\rm eff}$    &  $F(\nu)$  & $s_{0}$ \\
     &          &               &     (arcsec)   &  (mag/$\Box\arcsec$)   &   (mag)        &   (arcsec)    & (mag/$\Box\arcsec$)&  \\
\hline
2727 & IC4026  & 0.32 $\pm$ 0.01 &  -1.62 $\pm$ 0.04 & 15.19 $\pm$ 0.06 & 14.56 $\pm$ 0.24 &  5.57 $\pm$ 0.16 & 20.28 $\pm$ 0.09 &  9.46 $\pm$ 0.18 &  -6.88 \\
2736 &         & 0.55 $\pm$ 0.01 &  -0.53 $\pm$ 0.04 & 18.59 $\pm$ 0.07 & 17.07 $\pm$ 0.26 &  2.71 $\pm$ 0.08 & 21.22 $\pm$ 0.11 &  3.18 $\pm$ 0.21 &  -8.42 \\
2753 &         & 0.82 $\pm$ 0.01 &   0.06 $\pm$ 0.01 & 19.78 $\pm$ 0.03 & 17.02 $\pm$ 0.09 &  2.86 $\pm$ 0.03 & 21.29 $\pm$ 0.05 &  0.66 $\pm$ 0.08 &  -8.13 \\
2777 &         & 0.52 $\pm$ 0.00 &  -1.01 $\pm$ 0.01 & 16.69 $\pm$ 0.01 & 17.35 $\pm$ 0.03 &  1.05 $\pm$ 0.00 & 19.46 $\pm$ 0.01 &  3.51 $\pm$ 0.02 & -10.00 \\
2787 &         & 0.80 $\pm$ 0.02 &   0.06 $\pm$ 0.02 & 20.30 $\pm$ 0.03 & 17.46 $\pm$ 0.11 &  3.01 $\pm$ 0.04 & 21.85 $\pm$ 0.06 &  0.74 $\pm$ 0.10 &  -8.26 \\
2805 &         & 0.42 $\pm$ 0.02 &  -1.15 $\pm$ 0.11 & 15.78 $\pm$ 0.18 & 15.57 $\pm$ 0.72 &  2.37 $\pm$ 0.20 & 19.44 $\pm$ 0.29 &  5.67 $\pm$ 0.56 &  -8.19 \\
2839 & IC4021  & 0.37 $\pm$ 0.01 &  -1.49 $\pm$ 0.02 & 14.78 $\pm$ 0.03 & 14.90 $\pm$ 0.11 &  2.88 $\pm$ 0.04 & 19.19 $\pm$ 0.04 &  7.60 $\pm$ 0.08 &  -7.81 \\
2879 &         & 0.69 $\pm$ 0.02 &  -0.13 $\pm$ 0.03 & 19.42 $\pm$ 0.05 & 17.06 $\pm$ 0.17 &  2.82 $\pm$ 0.05 & 21.31 $\pm$ 0.08 &  1.49 $\pm$ 0.14 &  -8.22 \\
2897 &         & 0.62 $\pm$ 0.02 &  -0.17 $\pm$ 0.03 & 18.69 $\pm$ 0.06 & 16.00 $\pm$ 0.22 &  3.86 $\pm$ 0.10 & 20.93 $\pm$ 0.10 &  2.26 $\pm$ 0.19 &  -7.32 \\
2910 &         & 0.81 $\pm$ 0.03 &   0.05 $\pm$ 0.03 & 18.35 $\pm$ 0.05 & 15.59 $\pm$ 0.18 &  2.86 $\pm$ 0.06 & 19.87 $\pm$ 0.09 &  0.69 $\pm$ 0.15 &  -7.47 \\
2910 &         & 0.47 $\pm$ 0.01 &  -0.70 $\pm$ 0.04 & 16.55 $\pm$ 0.06 & 15.00 $\pm$ 0.23 &  3.49 $\pm$ 0.09 & 19.71 $\pm$ 0.09 &  4.43 $\pm$ 0.17 &  -7.22 \\
2921 & NGC4889 & 0.32 $\pm$ 0.01 &  -1.03 $\pm$ 0.01 & 15.41 $\pm$ 0.01 & 11.61 $\pm$ 0.04 & 25.12 $\pm$ 0.12 & 20.60 $\pm$ 0.02 &  9.72 $\pm$ 0.03 &  -3.29 \\
2922 & IC4012  & 0.50 $\pm$ 0.01 &  -0.76 $\pm$ 0.03 & 15.79 $\pm$ 0.05 & 14.86 $\pm$ 0.16 &  2.39 $\pm$ 0.05 & 18.75 $\pm$ 0.07 &  3.97 $\pm$ 0.12 &  -7.67 \\
2940 & IC4011  & 0.41 $\pm$ 0.01 &  -1.07 $\pm$ 0.02 & 16.03 $\pm$ 0.04 & 15.10 $\pm$ 0.14 &  3.60 $\pm$ 0.06 & 19.88 $\pm$ 0.05 &  6.15 $\pm$ 0.10 &  -7.40 \\
2960 &         & 0.54 $\pm$ 0.01 &  -0.39 $\pm$ 0.02 & 18.16 $\pm$ 0.03 & 15.89 $\pm$ 0.10 &  3.86 $\pm$ 0.04 & 20.82 $\pm$ 0.04 &  3.24 $\pm$ 0.08 &  -7.35 \\
2975 & NGC4886 & 0.28 $\pm$ 0.01 &  -2.16 $\pm$ 0.02 & 14.22 $\pm$ 0.03 & 14.02 $\pm$ 0.15 &  7.46 $\pm$ 0.13 & 20.38 $\pm$ 0.05 & 12.45 $\pm$ 0.11 &  -6.56 \\
3058 &         & 1.03 $\pm$ 0.02 &   0.31 $\pm$ 0.01 & 20.39 $\pm$ 0.03 & 16.88 $\pm$ 0.07 &  3.29 $\pm$ 0.03 & 21.46 $\pm$ 0.04 & -0.25 $\pm$ 0.06 &  -7.83 \\
3073 & NGC4883 & 0.41 $\pm$ 0.01 &  -0.88 $\pm$ 0.03 & 16.03 $\pm$ 0.05 & 14.33 $\pm$ 0.19 &  4.91 $\pm$ 0.11 & 19.78 $\pm$ 0.08 &  5.90 $\pm$ 0.15 &  -6.55 \\
3084 &         & 0.43 $\pm$ 0.01 &  -0.99 $\pm$ 0.03 & 16.26 $\pm$ 0.06 & 15.36 $\pm$ 0.21 &  3.19 $\pm$ 0.08 & 19.87 $\pm$ 0.08 &  5.57 $\pm$ 0.16 &  -7.64 \\
3113 &         & 0.59 $\pm$ 0.01 &  -0.34 $\pm$ 0.02 & 18.95 $\pm$ 0.04 & 16.86 $\pm$ 0.13 &  3.13 $\pm$ 0.05 & 21.34 $\pm$ 0.06 &  2.61 $\pm$ 0.10 &  -8.06 \\
3126 &         & 1.01 $\pm$ 0.02 &   0.22 $\pm$ 0.01 & 19.71 $\pm$ 0.03 & 16.63 $\pm$ 0.09 &  2.75 $\pm$ 0.03 & 20.82 $\pm$ 0.05 & -0.17 $\pm$ 0.07 &  -7.98 \\
3133 &         & 0.58 $\pm$ 0.02 &  -0.47 $\pm$ 0.05 & 17.96 $\pm$ 0.09 & 16.46 $\pm$ 0.33 &  2.44 $\pm$ 0.09 & 20.39 $\pm$ 0.15 &  2.71 $\pm$ 0.26 &  -8.26 \\
3170 & IC3998  & 0.38 $\pm$ 0.02 &  -1.14 $\pm$ 0.16 & 16.03 $\pm$ 0.23 & 14.89 $\pm$ 1.02 &  4.53 $\pm$ 0.52 & 20.16 $\pm$ 0.39 &  6.90 $\pm$ 0.81 &  -7.04 \\
3170 & IC3998  & 0.37 $\pm$ 0.01 &  -1.13 $\pm$ 0.05 & 16.07 $\pm$ 0.07 & 14.64 $\pm$ 0.31 &  5.52 $\pm$ 0.19 & 20.34 $\pm$ 0.12 &  7.26 $\pm$ 0.24 &  -6.67 \\
3201 & NGC4876 & 0.51 $\pm$ 0.01 &  -0.53 $\pm$ 0.03 & 16.68 $\pm$ 0.05 & 14.76 $\pm$ 0.17 &  3.61 $\pm$ 0.07 & 19.54 $\pm$ 0.07 &  3.72 $\pm$ 0.13 &  -6.99 \\
3205 &         & 0.81 $\pm$ 0.03 &   0.01 $\pm$ 0.03 & 19.38 $\pm$ 0.06 & 16.82 $\pm$ 0.19 &  2.61 $\pm$ 0.06 & 20.90 $\pm$ 0.10 &  0.68 $\pm$ 0.17 &  -8.18 \\
3206 &         & 0.60 $\pm$ 0.03 &  -0.32 $\pm$ 0.06 & 17.76 $\pm$ 0.11 & 15.71 $\pm$ 0.38 &  2.97 $\pm$ 0.13 & 20.07 $\pm$ 0.17 &  2.43 $\pm$ 0.32 &  -7.59 \\
3213 &         & 0.43 $\pm$ 0.01 &  -1.04 $\pm$ 0.01 & 15.97 $\pm$ 0.01 & 15.31 $\pm$ 0.04 &  2.84 $\pm$ 0.01 & 19.57 $\pm$ 0.02 &  5.54 $\pm$ 0.03 &  -7.78 \\
3222 &         & 0.46 $\pm$ 0.01 &  -1.10 $\pm$ 0.01 & 15.55 $\pm$ 0.01 & 15.78 $\pm$ 0.03 &  1.63 $\pm$ 0.00 & 18.84 $\pm$ 0.01 &  4.76 $\pm$ 0.02 &  -8.75 \\
3269 &         & 0.43 $\pm$ 0.03 &  -1.06 $\pm$ 0.15 & 16.11 $\pm$ 0.25 & 15.67 $\pm$ 0.97 &  2.49 $\pm$ 0.28 & 19.65 $\pm$ 0.39 &  5.38 $\pm$ 0.76 &  -8.13 \\
3291 &         & 0.81 $\pm$ 0.04 &  -0.07 $\pm$ 0.04 & 19.01 $\pm$ 0.07 & 16.87 $\pm$ 0.23 &  2.16 $\pm$ 0.06 & 20.54 $\pm$ 0.12 &  0.68 $\pm$ 0.20 &  -8.48 \\
3291 &         & 0.43 $\pm$ 0.03 &  -0.82 $\pm$ 0.13 & 17.66 $\pm$ 0.19 & 15.90 $\pm$ 0.88 &  4.69 $\pm$ 0.46 & 21.25 $\pm$ 0.35 &  5.52 $\pm$ 0.74 &  -7.31 \\
3292 &         & 1.16 $\pm$ 0.05 &   0.13 $\pm$ 0.03 & 19.91 $\pm$ 0.07 & 17.52 $\pm$ 0.16 &  1.81 $\pm$ 0.04 & 20.80 $\pm$ 0.09 & -0.60 $\pm$ 0.12 &  -9.02 \\
3296 & NGC4875 & 0.45 $\pm$ 0.02 &  -0.95 $\pm$ 0.09 & 15.71 $\pm$ 0.15 & 14.96 $\pm$ 0.58 &  2.71 $\pm$ 0.18 & 19.12 $\pm$ 0.23 &  5.05 $\pm$ 0.44 &  -7.64 \\
3302 &         & 0.50 $\pm$ 0.01 &  -0.70 $\pm$ 0.05 & 17.76 $\pm$ 0.08 & 16.54 $\pm$ 0.30 &  2.73 $\pm$ 0.09 & 20.72 $\pm$ 0.12 &  3.95 $\pm$ 0.23 &  -8.24 \\
3329 & NGC4874 & 0.38 $\pm$ 0.01 &  -0.47 $\pm$ 0.01 & 16.79 $\pm$ 0.01 & 12.32 $\pm$ 0.05 & 20.86 $\pm$ 0.12 & 20.91 $\pm$ 0.02 &  6.86 $\pm$ 0.04 &  -3.57 \\
3340 &         & 1.75 $\pm$ 0.12 &   0.23 $\pm$ 0.02 & 21.08 $\pm$ 0.05 & 18.61 $\pm$ 0.11 &  1.53 $\pm$ 0.03 & 21.53 $\pm$ 0.07 & -1.44 $\pm$ 0.10 &  -9.82 \\
3352 & NGC4872 & 0.33 $\pm$ 0.01 &  -1.74 $\pm$ 0.08 & 14.19 $\pm$ 0.12 & 14.36 $\pm$ 0.52 &  3.69 $\pm$ 0.22 & 19.19 $\pm$ 0.19 &  9.19 $\pm$ 0.40 &  -7.38 \\
3367 & NGC4873 & 0.47 $\pm$ 0.01 &  -0.66 $\pm$ 0.03 & 16.33 $\pm$ 0.05 & 14.57 $\pm$ 0.20 &  3.84 $\pm$ 0.09 & 19.49 $\pm$ 0.08 &  4.44 $\pm$ 0.15 &  -6.87 \\
3367 & NGC4873 & 0.79 $\pm$ 0.01 &  -0.02 $\pm$ 0.01 & 17.54 $\pm$ 0.02 & 15.04 $\pm$ 0.06 &  2.63 $\pm$ 0.02 & 19.14 $\pm$ 0.03 &  0.84 $\pm$ 0.04 &  -7.36 \\
3383 &         & 0.55 $\pm$ 0.02 &  -0.56 $\pm$ 0.06 & 18.60 $\pm$ 0.11 & 17.34 $\pm$ 0.38 &  2.35 $\pm$ 0.10 & 21.19 $\pm$ 0.17 &  3.07 $\pm$ 0.29 &  -8.75 \\
3400 & IC3973  & 0.41 $\pm$ 0.01 &  -1.24 $\pm$ 0.06 & 14.55 $\pm$ 0.10 & 14.57 $\pm$ 0.37 &  2.27 $\pm$ 0.10 & 18.35 $\pm$ 0.15 &  6.03 $\pm$ 0.27 &  -7.83 \\
3414 & NGC4871 & 0.30 $\pm$ 0.01 &  -1.92 $\pm$ 0.17 & 14.40 $\pm$ 0.23 & 14.34 $\pm$ 1.08 &  5.26 $\pm$ 0.64 & 19.94 $\pm$ 0.38 & 10.69 $\pm$ 0.84 &  -7.01 \\
3423 & IC3976  & 0.40 $\pm$ 0.02 &  -1.29 $\pm$ 0.13 & 14.69 $\pm$ 0.22 & 14.73 $\pm$ 0.82 &  2.39 $\pm$ 0.22 & 18.61 $\pm$ 0.32 &  6.35 $\pm$ 0.61 &  -7.87 \\
3439 &         & 0.34 $\pm$ 0.01 &  -1.45 $\pm$ 0.13 & 16.65 $\pm$ 0.17 & 15.54 $\pm$ 0.82 &  6.31 $\pm$ 0.57 & 21.54 $\pm$ 0.30 &  8.90 $\pm$ 0.66 &  -7.08 \\
3484 &         & 0.39 $\pm$ 0.02 &  -1.24 $\pm$ 0.14 & 15.81 $\pm$ 0.21 & 15.38 $\pm$ 0.87 &  3.13 $\pm$ 0.31 & 19.85 $\pm$ 0.33 &  6.64 $\pm$ 0.68 &  -7.79 \\
3486 &         & 0.64 $\pm$ 0.05 &  -0.51 $\pm$ 0.10 & 18.23 $\pm$ 0.21 & 17.43 $\pm$ 0.65 &  1.54 $\pm$ 0.12 & 20.36 $\pm$ 0.30 &  2.01 $\pm$ 0.48 &  -9.33 \\
3487 &         & 0.18 $\pm$ 0.01 &  -4.73 $\pm$ 0.12 & 11.69 $\pm$ 0.14 & 15.09 $\pm$ 0.76 &  8.48 $\pm$ 0.72 & 21.73 $\pm$ 0.23 & 24.33 $\pm$ 0.59 &  -8.27 \\
3510 & NGC4869 & 0.42 $\pm$ 0.01 &  -0.85 $\pm$ 0.03 & 15.79 $\pm$ 0.04 & 13.92 $\pm$ 0.19 &  5.28 $\pm$ 0.11 & 19.53 $\pm$ 0.07 &  5.89 $\pm$ 0.15 &  -6.25 \\
3522 &         & 0.46 $\pm$ 0.01 &  -0.97 $\pm$ 0.05 & 15.90 $\pm$ 0.09 & 15.51 $\pm$ 0.33 &  2.15 $\pm$ 0.08 & 19.17 $\pm$ 0.14 &  4.70 $\pm$ 0.25 &  -8.20 \\
3534 &         & 0.51 $\pm$ 0.02 &  -0.51 $\pm$ 0.07 & 18.10 $\pm$ 0.11 & 16.11 $\pm$ 0.46 &  3.74 $\pm$ 0.19 & 20.96 $\pm$ 0.19 &  3.72 $\pm$ 0.38 &  -7.55 \\
3554 &         & 0.64 $\pm$ 0.02 &  -0.32 $\pm$ 0.05 & 18.34 $\pm$ 0.10 & 16.55 $\pm$ 0.32 &  2.43 $\pm$ 0.09 & 20.48 $\pm$ 0.15 &  2.03 $\pm$ 0.25 &  -8.25 \\
3557 &         & 0.58 $\pm$ 0.01 &  -0.29 $\pm$ 0.02 & 17.82 $\pm$ 0.03 & 15.45 $\pm$ 0.11 &  3.62 $\pm$ 0.05 & 20.24 $\pm$ 0.05 &  2.66 $\pm$ 0.09 &  -7.20 \\
3561 & NGC4865 & 0.45 $\pm$ 0.01 &  -0.78 $\pm$ 0.01 & 15.30 $\pm$ 0.02 & 13.76 $\pm$ 0.06 &  3.81 $\pm$ 0.03 & 18.66 $\pm$ 0.03 &  4.95 $\pm$ 0.05 &  -6.57 \\
\hline
\end{tabular}

%% file: marquez.t2.tex
\tabcolsep=0.6\tabcolsep
\begin{tabular}{l l c c c c c c c r}
\hline
GMP  & Name     &     $\nu$     &      $\log a$  &  $-2.5 \log \Sigma_0$  &  $L_{\rm Tot}$ & $R_{\rm eff}$ & $\mu_{\rm eff}$    &  $F(\nu)$  & $s_{0}$ \\
     &          &               &     (arcsec)   &  (mag/$\Box\arcsec$)   &   (mag)        &   (arcsec)    & (mag/$\Box\arcsec$)&  \\
\hline
3564 &         & 0.20 $\pm$ 0.02 &  -3.14 $\pm$ 0.68 & 16.71 $\pm$ 0.64 & 14.75 $\pm$ 4.63 & 62.56 $\pm$30.82 & 25.73 $\pm$ 1.40 & 21.06 $\pm$ 3.94 &  -4.73 \\
3565 &         & 0.57 $\pm$ 0.03 &  -0.25 $\pm$ 0.08 & 19.86 $\pm$ 0.13 & 17.19 $\pm$ 0.51 &  4.27 $\pm$ 0.25 & 22.34 $\pm$ 0.23 &  2.81 $\pm$ 0.42 &  -7.77 \\
3639 & NGC4867 & 0.49 $\pm$ 0.01 &  -0.72 $\pm$ 0.05 & 15.78 $\pm$ 0.10 & 14.56 $\pm$ 0.34 &  2.81 $\pm$ 0.11 & 18.80 $\pm$ 0.14 &  4.10 $\pm$ 0.25 &  -7.30 \\
3656 &         & 0.19 $\pm$ 0.01 &  -3.97 $\pm$ 0.05 & 13.32 $\pm$ 0.06 & 14.03 $\pm$ 0.33 & 24.10 $\pm$ 0.87 & 22.93 $\pm$ 0.10 & 22.96 $\pm$ 0.26 &  -6.06 \\
3664 & NGC4864 & 0.48 $\pm$ 0.01 &  -0.54 $\pm$ 0.04 & 16.28 $\pm$ 0.06 & 13.97 $\pm$ 0.30 &  4.91 $\pm$ 0.16 & 19.42 $\pm$ 0.13 &  4.39 $\pm$ 0.28 &  -6.23 \\
3681 &         & 0.32 $\pm$ 0.02 &  -1.54 $\pm$ 0.24 & 17.34 $\pm$ 0.30 & 15.97 $\pm$ 1.60 &  8.37 $\pm$ 1.47 & 22.58 $\pm$ 0.57 &  9.86 $\pm$ 1.33 &  -6.97 \\
3707 &         & 0.66 $\pm$ 0.02 &  -0.23 $\pm$ 0.04 & 18.79 $\pm$ 0.07 & 16.74 $\pm$ 0.25 &  2.61 $\pm$ 0.07 & 20.81 $\pm$ 0.11 &  1.78 $\pm$ 0.20 &  -8.21 \\
3719 &         & 0.49 $\pm$ 0.02 &  -0.66 $\pm$ 0.09 & 18.75 $\pm$ 0.15 & 17.17 $\pm$ 0.58 &  3.38 $\pm$ 0.22 & 21.81 $\pm$ 0.24 &  4.19 $\pm$ 0.46 &  -8.24 \\
3733 & IC3960  & 0.36 $\pm$ 0.01 &  -1.42 $\pm$ 0.10 & 15.17 $\pm$ 0.16 & 14.63 $\pm$ 0.64 &  4.12 $\pm$ 0.30 & 19.70 $\pm$ 0.24 &  7.95 $\pm$ 0.49 &  -7.19 \\
3782 &         & 0.44 $\pm$ 0.02 &  -0.99 $\pm$ 0.11 & 16.23 $\pm$ 0.18 & 15.50 $\pm$ 0.68 &  2.83 $\pm$ 0.22 & 19.75 $\pm$ 0.27 &  5.34 $\pm$ 0.52 &  -7.86 \\
3792 & NGC4860 & 0.36 $\pm$ 0.01 &  -1.22 $\pm$ 0.02 & 14.92 $\pm$ 0.04 & 13.65 $\pm$ 0.15 &  5.45 $\pm$ 0.09 & 19.33 $\pm$ 0.06 &  7.62 $\pm$ 0.12 &  -6.28 \\
3794 &         & 0.52 $\pm$ 0.03 &  -0.79 $\pm$ 0.10 & 16.72 $\pm$ 0.18 & 16.19 $\pm$ 0.62 &  1.86 $\pm$ 0.13 & 19.53 $\pm$ 0.26 &  3.61 $\pm$ 0.46 &  -8.62 \\
3794 &         & 0.60 $\pm$ 0.02 &  -0.51 $\pm$ 0.06 & 17.31 $\pm$ 0.11 & 16.21 $\pm$ 0.35 &  1.91 $\pm$ 0.08 & 19.61 $\pm$ 0.16 &  2.41 $\pm$ 0.26 &  -8.49 \\
3851 &         & 0.56 $\pm$ 0.03 &  -0.43 $\pm$ 0.07 & 17.81 $\pm$ 0.12 & 15.97 $\pm$ 0.44 &  2.99 $\pm$ 0.15 & 20.35 $\pm$ 0.19 &  2.95 $\pm$ 0.35 &  -7.75 \\
3855 &         & 0.62 $\pm$ 0.01 &  -0.20 $\pm$ 0.03 & 19.54 $\pm$ 0.05 & 17.04 $\pm$ 0.18 &  3.50 $\pm$ 0.07 & 21.75 $\pm$ 0.08 &  2.20 $\pm$ 0.14 &  -7.94 \\
3914 &         & 0.55 $\pm$ 0.01 &  -0.71 $\pm$ 0.03 & 16.09 $\pm$ 0.06 & 15.52 $\pm$ 0.20 &  1.72 $\pm$ 0.04 & 18.69 $\pm$ 0.09 &  3.11 $\pm$ 0.15 &  -8.39 \\
4103 &         & 0.75 $\pm$ 0.01 &  -0.04 $\pm$ 0.01 & 19.29 $\pm$ 0.02 & 16.72 $\pm$ 0.07 &  2.85 $\pm$ 0.02 & 20.99 $\pm$ 0.03 &  1.08 $\pm$ 0.06 &  -8.02 \\
4129 &         & 0.79 $\pm$ 0.02 &  -0.08 $\pm$ 0.02 & 19.51 $\pm$ 0.05 & 17.35 $\pm$ 0.14 &  2.23 $\pm$ 0.04 & 21.09 $\pm$ 0.07 &  0.80 $\pm$ 0.11 &  -8.66 \\
4200 &         & 0.55 $\pm$ 0.01 &  -0.61 $\pm$ 0.03 & 16.88 $\pm$ 0.05 & 15.81 $\pm$ 0.17 &  2.17 $\pm$ 0.04 & 19.49 $\pm$ 0.07 &  3.11 $\pm$ 0.13 &  -8.17 \\
4230 &         & 0.26 $\pm$ 0.01 &  -2.53 $\pm$ 0.03 & 13.50 $\pm$ 0.04 & 13.94 $\pm$ 0.17 &  7.15 $\pm$ 0.14 & 20.21 $\pm$ 0.06 & 14.01 $\pm$ 0.13 &  -6.77 \\
\hline
\end{tabular}

%% file: marquez.t3.tex
\tabcolsep=0.6\tabcolsep
\begin{tabular}{l c c c c c c c r}
\hline
Id.  &     $\nu$     &      $\log a$  &  $-2.5 \log \Sigma_0$  &  $L_{\rm Tot}$ & $R_{\rm eff}$ & $\mu_{\rm eff}$    &  $F(\nu)$  & $s_{0}$ \\
     &               &     (arcsec)   &  (mag/$\Box\arcsec$)   &   (mag)        &   (arcsec)    & (mag/$\Box\arcsec$)&  \\
\hline
152 & 1.11 $\pm$ 0.16 & -0.15 $\pm$ 0.09 & 18.77 $\pm$ 0.22 & 17.72 $\pm$ 0.54 &  1.01 $\pm$ 0.09 & 19.72 $\pm$ 0.30 & -0.48 $\pm$ 0.41 &  -9.99 \\
156 & 0.76 $\pm$ 0.07 & -0.21 $\pm$ 0.09 & 19.60 $\pm$ 0.19 & 17.93 $\pm$ 0.58 &  1.87 $\pm$ 0.15 & 21.28 $\pm$ 0.28 &  1.01 $\pm$ 0.46 &  -9.21 \\
175 & 0.72 $\pm$ 0.06 & -0.32 $\pm$ 0.09 & 18.45 $\pm$ 0.17 & 17.18 $\pm$ 0.54 &  1.64 $\pm$ 0.12 & 20.24 $\pm$ 0.26 &  1.26 $\pm$ 0.42 &  -9.07 \\
179 & 0.89 $\pm$ 0.15 & -0.14 $\pm$ 0.13 & 19.34 $\pm$ 0.29 & 17.82 $\pm$ 0.82 &  1.48 $\pm$ 0.18 & 20.67 $\pm$ 0.43 &  0.26 $\pm$ 0.65 &  -9.47 \\
182 & 0.84 $\pm$ 0.01 &  0.12 $\pm$ 0.01 & 18.82 $\pm$ 0.02 & 15.79 $\pm$ 0.06 &  3.14 $\pm$ 0.02 & 20.27 $\pm$ 0.03 &  0.53 $\pm$ 0.05 &  -7.42 \\
197 & 0.56 $\pm$ 0.01 & -0.51 $\pm$ 0.02 & 17.15 $\pm$ 0.04 & 15.64 $\pm$ 0.15 &  2.61 $\pm$ 0.05 & 19.72 $\pm$ 0.06 &  3.03 $\pm$ 0.12 &  -7.81 \\
202 & 0.80 $\pm$ 0.08 & -0.18 $\pm$ 0.10 & 18.46 $\pm$ 0.22 & 16.81 $\pm$ 0.64 &  1.75 $\pm$ 0.16 & 20.02 $\pm$ 0.32 &  0.77 $\pm$ 0.50 &  -8.78 \\
208 & 0.32 $\pm$ 0.01 & -1.75 $\pm$ 0.11 & 16.66 $\pm$ 0.15 & 16.49 $\pm$ 0.70 &  4.69 $\pm$ 0.45 & 21.84 $\pm$ 0.25 &  9.69 $\pm$ 0.55 &  -8.06 \\
209 & 1.04 $\pm$ 0.02 &  0.18 $\pm$ 0.01 & 19.44 $\pm$ 0.02 & 16.64 $\pm$ 0.06 &  2.34 $\pm$ 0.02 & 20.49 $\pm$ 0.03 & -0.29 $\pm$ 0.05 &  -8.23 \\
212 & 0.89 $\pm$ 0.01 & -0.23 $\pm$ 0.01 & 19.03 $\pm$ 0.02 & 17.91 $\pm$ 0.04 &  1.24 $\pm$ 0.01 & 20.38 $\pm$ 0.02 &  0.30 $\pm$ 0.03 &  -9.77 \\
214 & 0.60 $\pm$ 0.01 & -0.23 $\pm$ 0.02 & 18.74 $\pm$ 0.03 & 16.18 $\pm$ 0.11 &  3.80 $\pm$ 0.06 & 21.08 $\pm$ 0.05 &  2.50 $\pm$ 0.09 &  -7.45 \\
215 & 0.84 $\pm$ 0.01 & -0.29 $\pm$ 0.01 & 18.78 $\pm$ 0.03 & 17.81 $\pm$ 0.08 &  1.22 $\pm$ 0.01 & 20.24 $\pm$ 0.04 &  0.55 $\pm$ 0.07 &  -9.76 \\
218 & 1.11 $\pm$ 0.07 &  0.10 $\pm$ 0.03 & 20.55 $\pm$ 0.08 & 18.24 $\pm$ 0.20 &  1.80 $\pm$ 0.05 & 21.51 $\pm$ 0.12 & -0.47 $\pm$ 0.17 &  -9.36 \\
221 & 0.28 $\pm$ 0.01 & -1.92 $\pm$ 0.04 & 15.74 $\pm$ 0.05 & 14.32 $\pm$ 0.27 & 13.18 $\pm$ 0.48 & 21.91 $\pm$ 0.09 & 12.47 $\pm$ 0.22 &  -5.84 \\
222 & 0.83 $\pm$ 0.05 & -0.23 $\pm$ 0.05 & 19.14 $\pm$ 0.11 & 17.82 $\pm$ 0.32 &  1.45 $\pm$ 0.07 & 20.62 $\pm$ 0.16 &  0.60 $\pm$ 0.25 &  -9.52 \\
225 & 0.67 $\pm$ 0.01 & -0.40 $\pm$ 0.01 & 18.52 $\pm$ 0.03 & 17.34 $\pm$ 0.07 &  1.71 $\pm$ 0.02 & 20.51 $\pm$ 0.04 &  1.70 $\pm$ 0.06 &  -9.11 \\
228 & 1.28 $\pm$ 0.44 &  0.09 $\pm$ 0.13 & 20.47 $\pm$ 0.33 & 18.44 $\pm$ 0.84 &  1.44 $\pm$ 0.43 & 21.23 $\pm$ 0.53 & -0.86 $\pm$ 0.80 &  -9.79 \\
229 & 0.88 $\pm$ 0.14 & -0.18 $\pm$ 0.12 & 20.16 $\pm$ 0.26 & 18.74 $\pm$ 0.78 &  1.44 $\pm$ 0.16 & 21.53 $\pm$ 0.41 &  0.36 $\pm$ 0.65 &  -9.94 \\
235 & 0.98 $\pm$ 0.13 & -0.02 $\pm$ 0.08 & 19.42 $\pm$ 0.18 & 17.49 $\pm$ 0.53 &  1.66 $\pm$ 0.13 & 20.58 $\pm$ 0.29 & -0.07 $\pm$ 0.46 &  -9.14 \\
236 & 0.55 $\pm$ 0.01 & -0.57 $\pm$ 0.02 & 16.97 $\pm$ 0.04 & 15.67 $\pm$ 0.12 &  2.43 $\pm$ 0.04 & 19.59 $\pm$ 0.06 &  3.16 $\pm$ 0.09 &  -7.94 \\
238 & 0.38 $\pm$ 0.01 & -1.49 $\pm$ 0.01 & 16.64 $\pm$ 0.02 & 17.17 $\pm$ 0.09 &  2.13 $\pm$ 0.03 & 20.81 $\pm$ 0.04 &  6.99 $\pm$ 0.07 &  -9.24 \\
242 & 0.98 $\pm$ 0.01 &  0.89 $\pm$ 0.00 & 20.23 $\pm$ 0.00 & 13.75 $\pm$ 0.01 & 13.41 $\pm$ 0.01 & 21.38 $\pm$ 0.00 & -0.08 $\pm$ 0.01 &  -4.28 \\
243 & 0.32 $\pm$ 0.01 & -1.78 $\pm$ 0.03 & 15.19 $\pm$ 0.04 & 15.17 $\pm$ 0.19 &  4.37 $\pm$ 0.12 & 20.37 $\pm$ 0.07 &  9.68 $\pm$ 0.15 &  -7.55 \\
246 & 1.75 $\pm$ 0.62 &  0.29 $\pm$ 0.05 & 22.77 $\pm$ 0.18 & 20.02 $\pm$ 0.40 &  1.74 $\pm$ 0.76 & 23.22 $\pm$ 0.33 & -1.44 $\pm$ 0.51 & -10.28 \\
253 & 0.76 $\pm$ 0.06 & -0.25 $\pm$ 0.08 & 18.75 $\pm$ 0.17 & 17.29 $\pm$ 0.51 &  1.68 $\pm$ 0.12 & 20.42 $\pm$ 0.25 &  1.01 $\pm$ 0.40 &  -9.07 \\
263 & 0.48 $\pm$ 0.01 & -1.06 $\pm$ 0.01 & 17.29 $\pm$ 0.02 & 17.67 $\pm$ 0.05 &  1.38 $\pm$ 0.01 & 20.37 $\pm$ 0.02 &  4.23 $\pm$ 0.04 &  -9.82 \\
283 & 0.64 $\pm$ 0.01 & -0.17 $\pm$ 0.02 & 18.90 $\pm$ 0.03 & 16.43 $\pm$ 0.11 &  3.29 $\pm$ 0.05 & 21.01 $\pm$ 0.05 &  1.96 $\pm$ 0.09 &  -7.73 \\
305 & 1.16 $\pm$ 0.06 &  0.11 $\pm$ 0.02 & 20.09 $\pm$ 0.06 & 17.80 $\pm$ 0.15 &  1.72 $\pm$ 0.04 & 20.98 $\pm$ 0.09 & -0.61 $\pm$ 0.13 &  -9.23 \\
316 & 1.11 $\pm$ 0.07 &  0.28 $\pm$ 0.03 & 19.92 $\pm$ 0.08 & 16.73 $\pm$ 0.21 &  2.70 $\pm$ 0.08 & 20.88 $\pm$ 0.12 & -0.48 $\pm$ 0.17 &  -8.05 \\
324 & 0.34 $\pm$ 0.01 & -1.43 $\pm$ 0.01 & 17.92 $\pm$ 0.02 & 16.93 $\pm$ 0.07 &  5.72 $\pm$ 0.05 & 22.71 $\pm$ 0.03 &  8.64 $\pm$ 0.06 &  -7.84 \\
326 & 0.70 $\pm$ 0.04 & -0.36 $\pm$ 0.06 & 18.56 $\pm$ 0.11 & 17.38 $\pm$ 0.36 &  1.62 $\pm$ 0.08 & 20.42 $\pm$ 0.17 &  1.42 $\pm$ 0.28 &  -9.19 \\
329 & 0.81 $\pm$ 0.02 & -0.25 $\pm$ 0.02 & 19.94 $\pm$ 0.04 & 18.69 $\pm$ 0.12 &  1.44 $\pm$ 0.02 & 21.48 $\pm$ 0.06 &  0.72 $\pm$ 0.10 &  -9.93 \\
413 & 1.07 $\pm$ 0.08 &  0.02 $\pm$ 0.04 & 19.94 $\pm$ 0.10 & 17.96 $\pm$ 0.28 &  1.59 $\pm$ 0.06 & 20.96 $\pm$ 0.16 & -0.36 $\pm$ 0.24 &  -9.41 \\
447 & 0.47 $\pm$ 0.01 & -1.00 $\pm$ 0.01 & 16.66 $\pm$ 0.01 & 16.61 $\pm$ 0.03 &  1.76 $\pm$ 0.01 & 19.83 $\pm$ 0.01 &  4.47 $\pm$ 0.03 &  -8.98 \\
\hline
\end{tabular}

%% file: marquez.t4.tex
\tabcolsep=0.6\tabcolsep
\begin{tabular}{l c c c c c c c r}
\hline
Id.  &     $\nu$     &      $\log a$  &  $-2.5 \log \Sigma_0$  &  $L_{\rm Tot}$ & $R_{\rm eff}$ & $\mu_{\rm eff}$    &  $F(\nu)$  & $s_{0}$ \\
     &               &     (arcsec)   &  (mag/$\Box\arcsec$)   &   (mag)        &   (arcsec)    & (mag/$\Box\arcsec$)&  \\
\hline
207 & 1.18 $\pm$ 0.18 &  0.03 $\pm$ 0.08 & 20.92 $\pm$ 0.19 & 19.06 $\pm$ 0.49 & 1.40 $\pm$ 0.12 & 21.78 $\pm$ 0.28 & -0.65 $\pm$ 0.41 & -10.12 \\
216 & 1.38 $\pm$ 0.08 &  0.37 $\pm$ 0.02 & 21.27 $\pm$ 0.04 & 17.89 $\pm$ 0.11 & 2.59 $\pm$ 0.04 & 21.95 $\pm$ 0.08 & -1.02 $\pm$ 0.12 &  -8.67 \\
237 & 0.35 $\pm$ 0.00 & -1.86 $\pm$ 0.00 & 14.36 $\pm$ 0.01 & 15.87 $\pm$ 0.02 & 1.68 $\pm$ 0.00 & 18.99 $\pm$ 0.01 &  8.19 $\pm$ 0.01 &  -9.13 \\
243 & 1.18 $\pm$ 0.12 &  0.15 $\pm$ 0.05 & 21.04 $\pm$ 0.11 & 18.56 $\pm$ 0.30 & 1.86 $\pm$ 0.09 & 21.91 $\pm$ 0.18 & -0.65 $\pm$ 0.28 &  -9.46 \\
243 & 0.92 $\pm$ 0.12 & -0.02 $\pm$ 0.09 & 20.43 $\pm$ 0.21 & 18.37 $\pm$ 0.60 & 1.86 $\pm$ 0.16 & 21.71 $\pm$ 0.32 &  0.16 $\pm$ 0.50 &  -9.38 \\
247 & 1.02 $\pm$ 0.07 &  0.16 $\pm$ 0.04 & 21.50 $\pm$ 0.08 & 18.75 $\pm$ 0.24 & 2.33 $\pm$ 0.08 & 22.58 $\pm$ 0.13 & -0.23 $\pm$ 0.22 &  -9.20 \\
254 & 0.41 $\pm$ 0.01 & -1.18 $\pm$ 0.07 & 15.65 $\pm$ 0.11 & 15.25 $\pm$ 0.42 & 2.83 $\pm$ 0.16 & 19.51 $\pm$ 0.17 &  6.18 $\pm$ 0.32 &  -7.84 \\
257 & 1.12 $\pm$ 0.05 &  0.18 $\pm$ 0.02 & 20.31 $\pm$ 0.05 & 17.60 $\pm$ 0.14 & 2.14 $\pm$ 0.04 & 21.25 $\pm$ 0.09 & -0.51 $\pm$ 0.14 &  -8.81 \\
258 & 1.10 $\pm$ 0.04 &  0.14 $\pm$ 0.02 & 20.13 $\pm$ 0.05 & 17.58 $\pm$ 0.13 & 2.02 $\pm$ 0.03 & 21.10 $\pm$ 0.07 & -0.45 $\pm$ 0.11 &  -8.89 \\
259 & 0.32 $\pm$ 0.00 & -1.90 $\pm$ 0.03 & 14.55 $\pm$ 0.05 & 15.00 $\pm$ 0.21 & 3.64 $\pm$ 0.11 & 19.80 $\pm$ 0.08 &  9.89 $\pm$ 0.16 &  -7.78 \\
260 & 0.89 $\pm$ 0.06 & -0.09 $\pm$ 0.05 & 19.48 $\pm$ 0.12 & 17.67 $\pm$ 0.33 & 1.68 $\pm$ 0.08 & 20.80 $\pm$ 0.17 &  0.26 $\pm$ 0.27 &  -9.21 \\
261 & 0.94 $\pm$ 0.07 & -0.08 $\pm$ 0.06 & 19.03 $\pm$ 0.13 & 17.32 $\pm$ 0.35 & 1.55 $\pm$ 0.08 & 20.27 $\pm$ 0.19 &  0.08 $\pm$ 0.28 &  -9.16 \\
262 & 0.83 $\pm$ 0.07 & -0.14 $\pm$ 0.07 & 19.36 $\pm$ 0.16 & 17.59 $\pm$ 0.46 & 1.79 $\pm$ 0.12 & 20.85 $\pm$ 0.23 &  0.60 $\pm$ 0.36 &  -9.10 \\
264 & 0.53 $\pm$ 0.01 & -0.67 $\pm$ 0.02 & 17.06 $\pm$ 0.04 & 16.11 $\pm$ 0.13 & 2.16 $\pm$ 0.04 & 19.77 $\pm$ 0.06 &  3.36 $\pm$ 0.10 &  -8.34 \\
266 & 0.83 $\pm$ 0.03 & -0.29 $\pm$ 0.04 & 17.97 $\pm$ 0.08 & 17.00 $\pm$ 0.23 & 1.23 $\pm$ 0.04 & 19.44 $\pm$ 0.12 &  0.58 $\pm$ 0.17 &  -9.39 \\
267 & 0.90 $\pm$ 0.07 & -0.12 $\pm$ 0.06 & 18.46 $\pm$ 0.15 & 16.84 $\pm$ 0.40 & 1.54 $\pm$ 0.09 & 19.77 $\pm$ 0.21 &  0.24 $\pm$ 0.31 &  -8.97 \\
268 & 0.83 $\pm$ 0.03 & -0.23 $\pm$ 0.03 & 18.95 $\pm$ 0.07 & 17.67 $\pm$ 0.20 & 1.41 $\pm$ 0.04 & 20.41 $\pm$ 0.10 &  0.56 $\pm$ 0.15 &  -9.49 \\
272 & 0.91 $\pm$ 0.07 &  0.03 $\pm$ 0.06 & 21.05 $\pm$ 0.12 & 18.71 $\pm$ 0.36 & 2.12 $\pm$ 0.11 & 22.34 $\pm$ 0.19 &  0.20 $\pm$ 0.30 &  -9.34 \\
287 & 0.55 $\pm$ 0.01 & -0.54 $\pm$ 0.03 & 17.38 $\pm$ 0.05 & 15.97 $\pm$ 0.19 & 2.52 $\pm$ 0.06 & 19.98 $\pm$ 0.08 &  3.08 $\pm$ 0.15 &  -8.02 \\
288 & 0.68 $\pm$ 0.01 & -0.32 $\pm$ 0.02 & 17.75 $\pm$ 0.04 & 16.27 $\pm$ 0.14 & 1.92 $\pm$ 0.04 & 19.69 $\pm$ 0.07 &  1.58 $\pm$ 0.11 &  -8.44 \\
291 & 0.41 $\pm$ 0.01 & -0.98 $\pm$ 0.03 & 16.08 $\pm$ 0.05 & 14.81 $\pm$ 0.20 & 4.07 $\pm$ 0.11 & 19.85 $\pm$ 0.08 &  5.97 $\pm$ 0.16 &  -7.06 \\
293 & 0.96 $\pm$ 0.03 &  0.03 $\pm$ 0.02 & 19.95 $\pm$ 0.05 & 17.74 $\pm$ 0.14 & 1.91 $\pm$ 0.04 & 21.13 $\pm$ 0.08 & -0.02 $\pm$ 0.12 &  -9.05 \\
294 & 0.47 $\pm$ 0.00 & -0.85 $\pm$ 0.02 & 16.07 $\pm$ 0.03 & 15.15 $\pm$ 0.10 & 2.70 $\pm$ 0.04 & 19.30 $\pm$ 0.04 &  4.60 $\pm$ 0.08 &  -7.69 \\
295 & 0.87 $\pm$ 0.04 & -0.03 $\pm$ 0.04 & 19.82 $\pm$ 0.09 & 17.64 $\pm$ 0.26 & 2.05 $\pm$ 0.07 & 21.20 $\pm$ 0.14 &  0.39 $\pm$ 0.22 &  -8.91 \\
304 & 1.00 $\pm$ 0.04 &  0.08 $\pm$ 0.03 & 19.75 $\pm$ 0.06 & 17.37 $\pm$ 0.17 & 2.00 $\pm$ 0.05 & 20.87 $\pm$ 0.09 & -0.14 $\pm$ 0.15 &  -8.80 \\
306 & 0.78 $\pm$ 0.01 & -0.15 $\pm$ 0.02 & 18.43 $\pm$ 0.04 & 16.58 $\pm$ 0.12 & 1.96 $\pm$ 0.03 & 20.03 $\pm$ 0.06 &  0.85 $\pm$ 0.09 &  -8.51 \\
309 & 0.44 $\pm$ 0.01 & -0.93 $\pm$ 0.08 & 16.48 $\pm$ 0.12 & 15.41 $\pm$ 0.49 & 3.33 $\pm$ 0.22 & 20.01 $\pm$ 0.19 &  5.37 $\pm$ 0.39 &  -7.57 \\
311 & 0.85 $\pm$ 0.03 &  0.00 $\pm$ 0.03 & 20.37 $\pm$ 0.06 & 18.00 $\pm$ 0.20 & 2.30 $\pm$ 0.06 & 21.81 $\pm$ 0.10 &  0.50 $\pm$ 0.17 &  -8.90 \\
313 & 0.97 $\pm$ 0.04 &  0.10 $\pm$ 0.03 & 20.55 $\pm$ 0.06 & 17.99 $\pm$ 0.18 & 2.22 $\pm$ 0.05 & 21.72 $\pm$ 0.10 & -0.05 $\pm$ 0.16 &  -8.93 \\
319 & 1.00 $\pm$ 0.13 & -0.03 $\pm$ 0.09 & 21.01 $\pm$ 0.20 & 19.15 $\pm$ 0.54 & 1.57 $\pm$ 0.13 & 22.13 $\pm$ 0.29 & -0.16 $\pm$ 0.44 &  -9.98 \\
322 & 0.50 $\pm$ 0.03 & -0.70 $\pm$ 0.12 & 18.16 $\pm$ 0.19 & 16.91 $\pm$ 0.74 & 2.80 $\pm$ 0.28 & 21.14 $\pm$ 0.31 &  4.00 $\pm$ 0.58 &  -8.38 \\
326 & 0.46 $\pm$ 0.01 & -0.93 $\pm$ 0.07 & 17.17 $\pm$ 0.11 & 16.54 $\pm$ 0.42 & 2.44 $\pm$ 0.14 & 20.48 $\pm$ 0.17 &  4.79 $\pm$ 0.33 &  -8.50 \\
331 & 0.82 $\pm$ 0.02 & -0.04 $\pm$ 0.02 & 18.76 $\pm$ 0.04 & 16.49 $\pm$ 0.14 & 2.26 $\pm$ 0.04 & 20.25 $\pm$ 0.07 &  0.61 $\pm$ 0.12 &  -8.24 \\
333 & 0.41 $\pm$ 0.00 & -0.86 $\pm$ 0.01 & 16.58 $\pm$ 0.02 & 14.63 $\pm$ 0.07 & 5.66 $\pm$ 0.06 & 20.40 $\pm$ 0.03 &  6.09 $\pm$ 0.06 &  -6.50 \\
\hline
\end{tabular}

%% file: marquez.bbl
\begin{thebibliography}{}

\bibitem{Bellazzini} Bellazzini M., 1998, New Astronomy 4, 219
\bibitem{Binggeli} Binggeli B., Jerjen H., 1998, A\&A 333, 17
\bibitem{Binney} Binney J., Tremaine S., 1987, ``Galactic Dynamics'',
Princeton Series in Astrophysics, Princeton, New Jersey
\bibitem{Biviano} Biviano A., Durret F., Gerbal D., et al., 1995, 
A\&AS 111, 265
\bibitem{Caon} Caon N., Capaccioli M., D'Onofrio M., 1993, MNRAS 265, 1013
\bibitem{Capelato} Capelato H.  V., de Carvalho R.R., Carlberg R.G., 
1995, ApJ 451, 525
\bibitem{Capelato2} Capelato H.  V., de Carvalho R.R., Carlberg R.G., 
1997, in: `Galaxy scaling relations: Origins, evolution and application', p.~331,
L.N. da Costa,  A. Renzini Editors, Spring-Verlag
\bibitem{Durret} Durret F., Felenbok P., Lobo C., Slezak E., 1998, 
A\&AS 129, 281
\bibitem{Durret1} Durret F., Felenbok P., Lobo C., Slezak E., 1999, 
A\&AS in press, astro-ph/9908045
\bibitem{Fujita} Fujita Y., Takahara F., 1999, ApJ Letters in press, astro-ph/9905082
\bibitem{GMP} Godwin J., Metcalfe N., Peach J.V., 1983, MNRAS 202, 113
\bibitem{Graham} Graham A., Colless M., 1997, MNRAS 287, 221
\bibitem{Guzman} Guzm\'an R., Lucey J.R., Bower R.G.R.A.S., 1993, MNRAS 265, 731
\bibitem{Hernquist} Hernquist L., 1990, ApJ 356, 359
\bibitem{Lima} Lima Neto G.B., Gerbal D., M\'arquez I., 1999a, MNRAS 309, 481
\bibitem{Lima2} Lima Neto G.B., Gerbal D., M\'arquez I., 1999b, Proceedings of the
XIV Moriond Meeting, in press
\bibitem{Lobo} Lobo C., Biviano A., Durret F., et al., 1997, A\&AS 122, 409
\bibitem{Lynden-Bell} Lynden-Bell D., 1967, MNRAS 136, 101
\bibitem{Markevitch} Markevitch M., 1998, ApJ 504, 27
\bibitem{Merritt} Merritt D., 1999, PASP 111,129
\bibitem{Mo} Mo H.J., Mao S., White S.D.M., 1998, MNRAS 295, 319
\bibitem{Navarro} Navarro J.F., Frenk C.S., White S.D.M., 1995, MNRAS 275, 720
\bibitem{Prugniel} Prugniel P., Simien F., 1997, A\&A 321, 111
\bibitem{saslaw} Saslaw W.C., 1985, `Gravitational Physics of Stellar and
Galactic Systems', Cambridge Univ. Press
\bibitem{Slezak} Slezak E., Durret F., Guibert J., Lobo C., 1998, A\&AS 128, 67
\bibitem{Slezak1} Slezak E., Durret F., Guibert J., Lobo C., 1998, 
A\&AS in press, astro-ph/9908068
\bibitem{Tormen} Tormen G., Bouchet F.R., White S.D.M., 1997, MNRAS 286, 865
\bibitem{Tremaine} Tremaine S., H\'enon M., Lynden-Bell D., 1986, MNRAS 219, 285
\bibitem{Wiechen} Wiechen H, Ziegler H.J., Schindler K, 1988, MNRAS 232, 623
\bibitem{White} White S.D.M., Narayan, R., 1987, MNRAS 229, 103
\bibitem{Young} Young C. K., Currie M. J., 1994, MNRAS 268, L11
\bibitem{Young1} Young C. K., Currie M. J., 1995, MNRAS 273, 1141
\end{thebibliography}
